\documentclass[12pt]{article}
\usepackage{graphicx}
\usepackage{lscape}
\usepackage{citesort}
\usepackage{amssymb}
\usepackage{appendix}
\usepackage{multirow}

\begin{document}
\def\qq{\langle \bar q q \rangle}
\def\uu{\langle \bar u u \rangle}
\def\dd{\langle \bar d d \rangle}
\def\sp{\langle \bar s s \rangle}
\def\GG{\langle g_s^2 G^2 \rangle}
\def\Tr{\mbox{Tr}}
\def\figt#1#2#3{
        \begin{figure}
        $\left. \right.$
        \vspace*{-2cm}
        \begin{center}
        \includegraphics[width=10cm]{#1}
        \end{center}
        \vspace*{-0.2cm}
        \caption{#3}
        \label{#2}
        \end{figure}
    }

\def\figb#1#2#3{
        \begin{figure}
        $\left. \right.$
        \vspace*{-1cm}
        \begin{center}
        \includegraphics[width=10cm]{#1}
        \end{center}
        \vspace*{-0.2cm}
        \caption{#3}
        \label{#2}
        \end{figure}
                }

\def\ds{\displaystyle}
\def\beq{\begin{equation}}
\def\eeq{\end{equation}}
\def\bea{\begin{eqnarray}}
\def\eea{\end{eqnarray}}
\def\beeq{\begin{eqnarray}}
\def\eeeq{\end{eqnarray}}
\def\ve{\vert}
\def\vel{\left|}
\def\ver{\right|}
\def\nnb{\nonumber}
\def\ga{\left(}
\def\dr{\right)}
\def\aga{\left\{}
\def\adr{\right\}}
\def\lla{\left<}
\def\rra{\right>}
\def\rar{\rightarrow}
\def\lrar{\leftrightarrow}
\def\nnb{\nonumber}
\def\la{\langle}
\def\ra{\rangle}
\def\ba{\begin{array}}
\def\ea{\end{array}}
\def\tr{\mbox{Tr}}
\def\ssp{{\Sigma^{*+}}}
\def\sso{{\Sigma^{*0}}}
\def\ssm{{\Sigma^{*-}}}
\def\xis0{{\Xi^{*0}}}
\def\xism{{\Xi^{*-}}}
\def\qs{\la \bar s s \ra}
\def\qu{\la \bar u u \ra}
\def\qd{\la \bar d d \ra}
\def\qq{\la \bar q q \ra}
\def\gGgG{\la g^2 G^2 \ra}
\def\q{\gamma_5 \not\!q}
\def\x{\gamma_5 \not\!x}
\def\g5{\gamma_5}
\def\sb{S_Q^{cf}}
\def\sd{S_d^{be}}
\def\su{S_u^{ad}}
\def\sbp{{S}_Q^{'cf}}
\def\sdp{{S}_d^{'be}}
\def\sup{{S}_u^{'ad}}
\def\ssp{{S}_s^{'??}}

\def\sig{\sigma_{\mu \nu} \gamma_5 p^\mu q^\nu}
\def\fo{f_0(\frac{s_0}{M^2})}
\def\ffi{f_1(\frac{s_0}{M^2})}
\def\fii{f_2(\frac{s_0}{M^2})}
\def\O{{\cal O}}
\def\sl{{\Sigma^0 \Lambda}}
\def\es{\!\!\! &=& \!\!\!}
\def\ap{\!\!\! &\approx& \!\!\!}
\def\md{\!\!\!\! &\mid& \!\!\!\!}
\def\ar{&+& \!\!\!}
\def\ek{&-& \!\!\!}
\def\kek{\!\!\!&-& \!\!\!}
\def\cp{&\times& \!\!\!}
\def\se{\!\!\! &\simeq& \!\!\!}
\def\eqv{&\equiv& \!\!\!}
\def\kpm{&\pm& \!\!\!}
\def\kmp{&\mp& \!\!\!}
\def\mcdot{\!\cdot\!}
\def\erar{&\rightarrow&}

% .........................................................

\def\simlt{\stackrel{<}{{}_\sim}}
\def\simgt{\stackrel{>}{{}_\sim}}

% .........................................................

\title{
         {\Large
                 {\bf
                     Light cone QCD sum rules study of the semileptonic heavy  $\Xi_{Q}$ and $\Xi'_{Q}$  transitions to  $\Xi$ and $\Sigma $ baryons
                 }
         }
      }

\author{\vspace{1cm}\\
{\small K. Azizi$^a$ \thanks {e-mail: kazizi@dogus.edu.tr}\,, Y.
Sarac$^b$
\thanks {e-mail: ysoymak@atilim.edu.tr}\,\,, H.
Sundu$^c$ \thanks {e-mail: hayriye.sundu@kocaeli.edu.tr}} \\
{\small $^a$  Physics Department, Do\u gu\c s University, Ac{\i}badem-Kad{\i}k\"oy, 34722 Istanbul, Turkey} \\
{\small $^b$ Electrical and Electronics Engineering Department,
Atilim University, 06836 Ankara, Turkey} \\
{\small $^c$ Department of Physics, Kocaeli University, 41380 Izmit,
Turkey}}
\date{}

\begin{titlepage}
\maketitle
\thispagestyle{empty}

\begin{abstract}
The semileptonic decays of heavy spin--1/2, $\Xi_{b(c)}$ and $\Xi'_{b(c)}$ baryons to the light spin-- 1/2, $\Xi$ and $\Sigma $ baryons
are investigated in the framework of light cone QCD sum rules. In particular, using the most general form of the interpolating currents for the heavy baryons as well as the distribution amplitudes of the
 $\Xi$ and $\Sigma $ baryons, we calculate all form factors entering the matrix elements of the corresponding effective Hamiltonians in full QCD. Having calculated the responsible form factors, we evaluate
 the decay rates and branching fractions
of the related transitions.

\end{abstract}

~~~PACS number(s): 11.55.Hx, 13.30.Ce, 14.20.Mr, 14.20.Lq
\end{titlepage}

%%%
\section{Introduction}

Almost all of the  anti-triplet states $\Lambda_c^{+},~\Xi_c^{+},~\Xi_c^{0}$ [ $\Lambda_c^{+} (2593)$,
$\Xi_c^{+}(2790),~\Xi_c^{0}(2790)$] with  $J^P={1\over 2}^+$ [ ${1\over 2}^-$] and containing single heavy charm quark as well as the ${1\over 2}^+$ [  ${3\over 2}^+$]  sextet
 $\Omega_c,\Sigma_c,\Xi'_c$ [$\Omega_c^\ast,\Sigma_c^\ast,\Xi_c^\ast$] states have been detected in the experiments
\cite{Rstp01}. Among the S--wave bottom baryons, the
$\Lambda_b,~\Sigma_b,~\Sigma_b^\ast,~\Xi_b$ and $\Omega_b$ states
have also  been observed. It is expected that the LHC not only will
open new horizons in the discovery of the excited bottom baryons but
also it will provide possibility to study  properties of heavy
baryons as well as  their electromagnetic, weak and strong decays.

Such an  experimental progress stimulates  the theoretical
studies on properties of  the heavy baryons as well as their electromagnetic, weak and strong transitions.  The mass spectrum of the heavy baryons has been
studied using various methods including heavy quark effective
theory \cite{Grozin}, QCD sum rules \cite{Navarra,Shuryak,Gang Wang,boyuk}
and some other phenomenological models
\cite{Karliner1,Karliner2,Ebert,Mathur,Rosner,Qiao-Yan Zhao}. Some electromagnetic properties of the heavy baryons and their radiative decays have been investigated 
in different frameworks in \cite{boyuk,ek1,ek2,ek3,ek4,ek5,ek6,ek7,ek8,ek9,ek10,ek11,T. M. Aliev}. The strong
decays of the heavy baryons have also been in the focus of much attention, theoretically (see for instance \cite{boyuk2,boyuk3,boyuk4} and references therein).

However, the weak and semileptonic decays of heavy baryons are very
important frameworks  not only in obtaining information about their
internal structure, precise calculation of the main ingredients of
standard model (SM) such as Kabbibbo-Kobayashi-Maskawa (CKM) matrix
elements and answering to some fundamental questions like nature of
the CP violation, but also in looking for new physics beyond the SM.
The loop level semileptonic transitions of the heavy baryons
containing single heavy quark to light baryons induced by the flavor
changing neutral currents (FCNC)
 are useful tools, for instance, to look for the supersymmetric particles, light dark matter, fourth generation of the quarks and extra dimensions etc. \cite{ek12,ek13}.
 Some semileptonic decay channels of the heavy baryons   have been
previously investigated in different frameworks (see for instance \cite{ebert3,Ming-Qiu,Albertus,Flores-Mendieta,Pervin,Azizi1,Azizi2,Azizi3,Azizi4} and  references therein).

The present work deals with the semileptonic decays of heavy
$\Xi_{b(c)}$ and $\Xi'_{b(c)}$ baryons to the light $\Xi$ and
$\Sigma$ baryons. The considered channels are either at
loop level described by twelve form factors in full QCD  or at tree
level analyzed by six form factors entering the transition matrix
elements of the corresponding low energy  Hamiltonian. Here, we should mention that by the  ``full QCD`` we refer to the QCD theory without any approximation like heavy quark effective theory (HQET)
 so we take the mass of heavy quarks finite. In HQET approximation, the number of form factors describing the considered transitions reduce to only two form factors \cite{menel,savci}. The considered
processes take place in low energies far from the perturbative
region, so to calculate the form factors as the main ingredients, we
should consult some nonperturbative methods. One of the most
powerful, applicable and attractive nonperturbative methods is QCD
sum rules \cite{shif1,shif2} and its extension light cone sum rules (LCSR) (see for instance \cite{balta}). We apply
the LCSR method to calculate the corresponding form factors in full
theory. In this approach, the time ordering multiplication of the
most general form of the interpolating currents for considered heavy
baryons with transition currents are expanded in terms of the
distribution amplitudes (DA's) of the  light $\Xi$ and $\Sigma$
baryons. Using the obtained form factors, we calculate the decay
rate and branching ratio for the considered channels.

The introduction is followed by  section 2 which
presents the details of the application of the LCSR method
to find the QCD sum rules for the form factors. Section 3 is devoted
to the numerical analysis of the form factors as well as evaluation of the   decay widths and  branching fractions. Finally, section 4 encompasses our conclusion.

%%%
\section{ LCSR for transition form factors}
This section is dedicated to the details of calculations of the form
factors. As we previously mentioned, the considered transitions can be classified as  loop FCNC  and tree level decays. The loop level
 transitions include the semileptonic
$\Xi_b\rightarrow\Xi l^+l^-$, $\Xi_b\rightarrow\Sigma l^+l^-$, $\Xi_c\rightarrow\Sigma l^+l^-$
$\Xi'_b\rightarrow\Xi l^+l^-$, $\Xi'_b\rightarrow\Sigma l^+l^-$ and
$\Xi'_c\rightarrow\Sigma l^+l^-$   decays. Considering the quark contents and charges of the participant baryons,  these channels  proceed
via FCNC $b\rightarrow s$, $b\rightarrow d$ or $c\rightarrow u$ transitions at quark
level. The low energy effective Hamiltonian describing the above transitions is written as:
\begin{eqnarray} \label{ham} {\cal H}^{loop}_{eff} \es \frac{G_F~\alpha_{em}
V_{Q'Q}~V_{Q'q}^{^{*}}}{2\sqrt2~\pi} \Bigg\{\vphantom{\int_0^{x_2}}
C_{9}^{eff}~ \bar{q} \gamma_\mu (1-\gamma_5) Q \bar l \gamma^\mu
 l +C_{10} ~\bar{q}
\gamma_\mu (1-\gamma_5) Q \bar l \gamma^\mu \gamma_{5}l \nnb \\
\ek2 m_{Q}~C_{7}^{eff}\frac{1}{q^{2}} ~\bar{q} i \sigma_{\mu\nu}q^{\nu}
(1+\gamma_5) Q \bar l \gamma^\mu l \Bigg\}~,
\end{eqnarray}
where $Q$ corresponds to $b$ or $c$ quark, $Q'$ represents the $t$
or $b$ quark and  $q$ denotes the $s$, $d$ or $u$ quark with
respect to the transition under consideration. The tree level
transitions include the channels, $\Xi_c\rightarrow \Xi l\nu$,
$\Xi_c\rightarrow \Sigma l\nu$, $\Xi'_c\rightarrow \Xi l\nu$ and
$\Xi'_c\rightarrow \Sigma l\nu$, which
 proceed via $c\rightarrow s$ or $c\rightarrow d$ depending on the quark contents and charges of the initial and final baryons.
The effective Hamiltonian representing the considered tree level
transitions has the following form:
\begin{eqnarray}\label{Eq1} {\cal H}^{tree}_{eff} = \frac{G_F}{\sqrt2}
V_{qc} ~\bar q \gamma_\mu (1-\gamma_5) c \bar l \gamma^\mu
(1-\gamma_5) \nu,
\end{eqnarray}
where $q$ can be either $s$ or $d$ quark, 
$G_F$ is the Fermi coupling constant, and $V_{Q'Q}$ , $V_{Q'q}$ and  $V_{qc}$ are elements of the  CKM matrix.

In order to get the amplitudes, we need to sandwich the effective Hamiltonians between the initial and final states. Looking at the effective Hamiltonians, 
we see that we have two transition currents, $J_\mu^{tr,I}=\bar q \gamma_\mu
(1-\gamma_5) Q$ and $ J_\mu^{tr,II}=\bar q i
\sigma_{\mu\nu}q^{\nu} (1- \gamma_5) Q$. The matrix elements of the transition currents are parameterized in terms of form factors in the following way:
\begin{eqnarray}\label{matrixel1a} \langle B(p) \md  J_\mu^{tr,I} \mid B_Q(p+q,s)   \rangle= \bar
{u}_{B}(p) \Big[\gamma_{\mu}f_{1}(Q^{2})+{i}
\sigma_{\mu\nu}q^{\nu}f_{2}(Q^{2})
+ q^{\mu}f_{3}(Q^{2})\nnb \\\ek \gamma_{\mu}\gamma_5
g_{1}(Q^{2})-{i}\sigma_{\mu\nu}\gamma_5q^{\nu}g_{2}(Q^{2}) -
q^{\mu}\gamma_5 g_{3}(Q^{2}) \vphantom{\int_0^{x_2}}\Big] u_{B_Q}(p+q,s)~,
\end{eqnarray}
and
\begin{eqnarray}\label{matrixel1b} \langle B(p)\md J_\mu^{tr,II} \mid B_Q(p+q,s) \rangle
=\bar{u}_{B}(p)
\Big[\gamma_{\mu}f_{1}^{T}(Q^{2})+{i}\sigma_{\mu\nu}q^{\nu}f_{2}^{T}(Q^{2})
+ q^{\mu}f_{3}^{T}(Q^{2})\nnb \\ \ar \gamma_{\mu}\gamma_5
g_{1}^{T}(Q^{2})+{i}\sigma_{\mu\nu}\gamma_5q^{\nu}g_{2}^{T}(Q^{2}) +
q^{\mu}\gamma_5 g_{3}^{T}(Q^{2}) \vphantom{\int_0^{x_2}}\Big]
u_{B_Q}(p+q,s)~,
\end{eqnarray}
where $Q^2=-q^2$, $f_i$, $g_i$,   $f^T_i$ and $g^T_i$ are transition
form factors, and $u_{B_Q}$ and   $u_{B}$ are 
spinors of the initial and final states. The $B_Q(p+q,s)$ stands for particles with momentum $p+q$ and spin $s$. From the explicit expressions of the
effective Hamiltonians,
 it is clear that the loop level transitions contain both transition matrix elements having twelve form factors
 while the tree level channels include only the transition current I that corresponds to six form factors.

\begin{figure}[h!]
\begin{center}
\includegraphics[width=6cm]{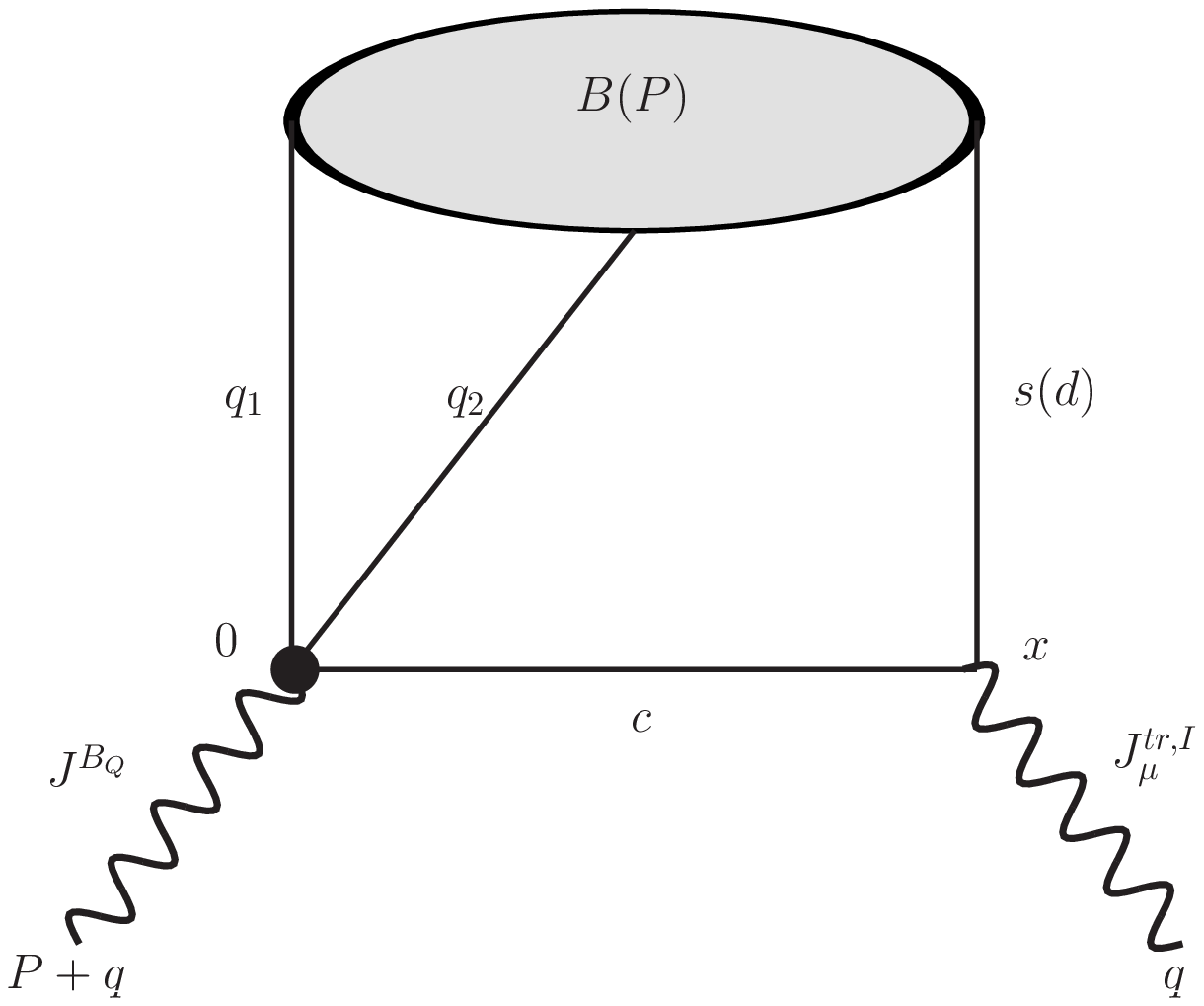}
\includegraphics[width=6cm]{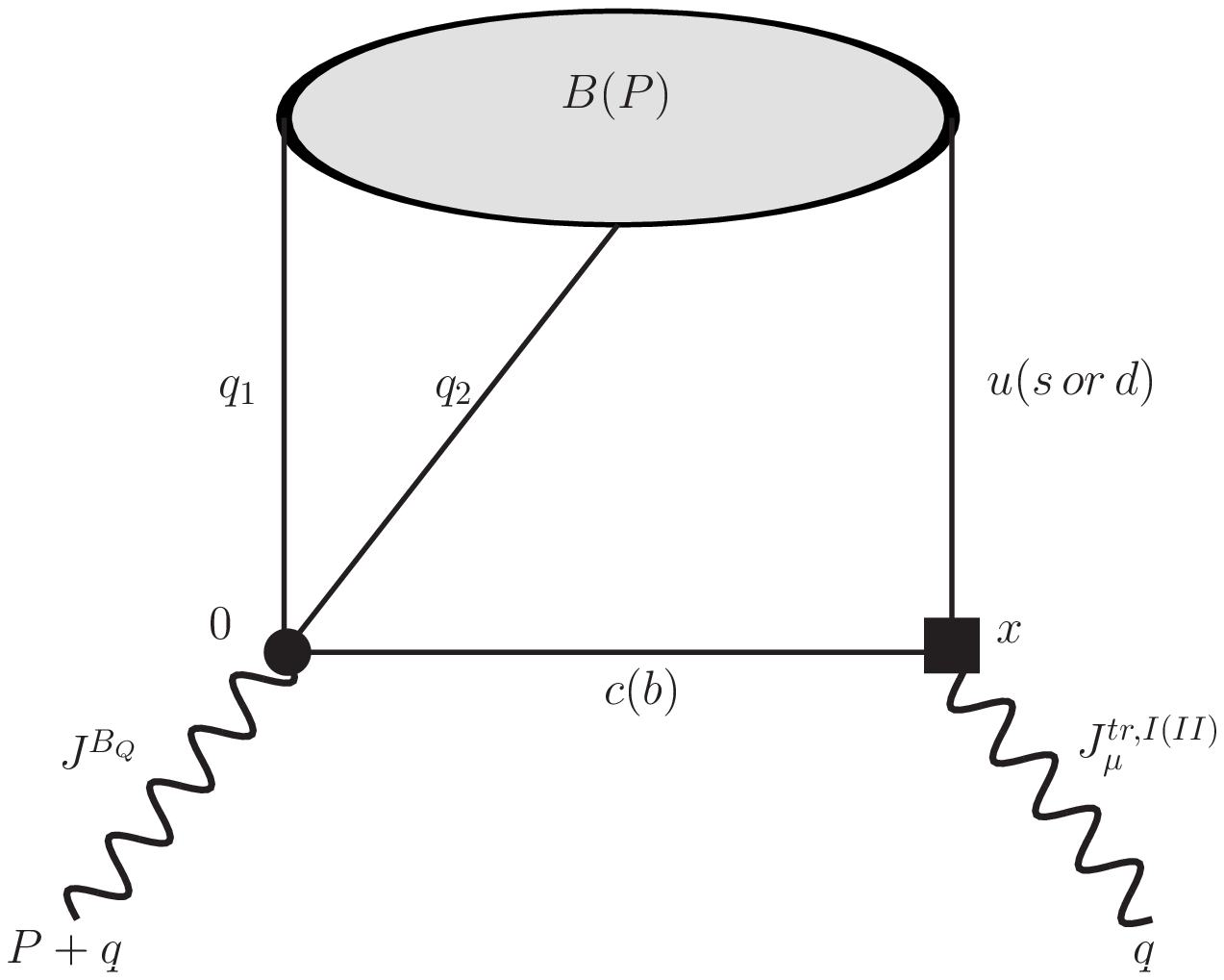}
\end{center}
\caption{Diagrammatic representations of  the correlation
functions given in Eq.~(\ref{T}). The left (right) diagram belongs to the tree (loop) level transitions. The ovals (wavy
lines) in  diagrams stand for the DA's of $\Xi$
or $\Sigma$ baryons (external currents). In each graph, the $q_1$ and $q_2$ are light spectator quarks.} \label{CorrelationFunc}
\end{figure}

Our main task in the present work is to calculate the transition form factors.
According to the philosophy of the QCD sum rules approach, we start with the following correlation functions as the main building blocks of the method:
\begin{eqnarray}\label{T}
\Pi^I_{\mu}(p,q) = i\int d^{4}xe^{iqx}\langle B(p) \mid
T\{ J_\mu^{tr,I}(x), \bar J^{B_Q}(0)\}\mid
0\rangle~,\nnb\\
\Pi^{II}_{\mu}(p,q) = i\int d^{4}xe^{iqx}\langle B(p) \mid
T\{ J_\mu^{tr,II}(x), \bar J^{B_Q}(0)\}\mid
0\rangle~,
\end{eqnarray}
 where  $J^{B_{Q}}$ is the
interpolating current carrying the quantum numbers of the
$\Xi_{Q}(\Xi'_Q)$ baryons. The diagrammatic representations
of these correlation functions are presented in
Figure~\ref{CorrelationFunc}. The interpolating currents for the
considered baryons
 have the following general forms (see for instance \cite{Bagan}):
\begin{eqnarray}\label{IPF}
J^{\Xi'_Q} \es - {1\over \sqrt{2}} \epsilon^{abc} \Big\{ \Big(
q_1^{aT} C Q^b \Big) \gamma_5 q_2^c + \beta \Big( q_1^{aT} C
\gamma_5 Q^b \Big) q_2^c - \Big[\Big( Q^{aT} C q_2^b \Big) \gamma_5
q_1^c + \beta \Big( Q^{aT} C
\gamma_5 q_2^b \Big) q_1^c \Big] \Big\}~, \nnb\\
\label{estp06} J^{\Xi_Q} \es {1\over \sqrt{6}} \epsilon^{abc} \Big\{
2 \Big( q_1^{aT} C q_2^b \Big) \gamma_5 Q^c + 2 \beta \Big( q_1^{aT}
C \gamma_5 q_2^b \Big) Q^c + \Big( q_1^{aT} C Q^b \Big) \gamma_5
q_2^c + \beta \Big(q_1^{aT} C
\gamma_5 Q^b \Big) q_2^c \nnb \\
\ar \Big(Q^{aT} C q_2^b \Big) \gamma_5 q_1^c + \beta \Big(Q^{aT} C
\gamma_5 q_2^b \Big) q_1^c \Big\}~,
\end{eqnarray}
where $C$ is the charge conjugation operator,  $a$,
$b$ and $c$ are color indices
and the light quarks $q_1$ and $q_2$ are given in Table~\ref{tab:1a}. The $\beta$ is an
arbitrary parameter and the value $\beta=-1$ corresponds to the Ioffe
current.

\begin{table}[h] \centering
\begin{tabular}{|l||c|c|c|c|c|} \hline &
$\Xi^{-(0)}_{b(c)}$ & $\Xi^{0(+)}_{b(c)}$& $\Xi'^{-(0)}_{b(c)}$ & $\Xi'^{0(+)}_{b(c)}$ \\
\hline\hline $q_1$ & d & u&d&u
\\ \hline
$q_2$ & s & s&s&s
\\ \hline
\end{tabular}
\vspace{0.8cm} \caption{The light quark contents of the heavy baryons
$\Xi_Q$ and $\Xi_Q'$.} \label{tab:1a}
\end{table}

The correlation functions given above can be calculated in two different ways. From the phenomenological or physical side, they are calculated  inserting
 complete sets of hadronic states having the same quantum numbers
as the chosen interpolating fields. The results
of this side appear in terms of hadronic degrees of freedom. On
the other side, the QCD or theoretical side of the correlation functions are calculated in terms of the $B$ baryon
DA's via operator product expansion (OPE). Then, we match these two different representations  to relate the hadronic parameters to fundamental QCD degrees of freedom which leads to  QCD sum rules for the considered form factors. To suppress  contribution of the higher states and continuum,
we apply Borel transformation with respect to the initial momentum squared to both sides of the sum rules and use the quark-hadron duality assumption.

Inserting  complete set of hadronic state into correlation functions
and isolating the contribution of the ground state, we
obtain the following representations from physical side:
\begin{eqnarray} \label{phys1} \Pi_{\mu}^{I}(p,q)&=&\sum_{s}\frac{\langle
B(p)\mid J^{tr,I}_\mu
\mid B_Q(p+q,s)\rangle\langle B_Q(p+q,s)\mid
\bar J^{B_Q}(0) \mid
0\rangle}{m_{B_Q}^{2}-(p+q)^{2}} +\cdots~,
\end{eqnarray}
\begin{eqnarray}
\label{phys1111} \Pi_{\mu}^{II}(p,q)&=&\sum_{s}\frac{\langle
B(p)\mid J^{tr,II}_\mu
\mid B_Q(p+q,s)\rangle\langle B_Q(p+q,s)\mid
\bar J^{B_Q}(0) \mid
0\rangle}{m_{B_Q}^{2}-(p+q)^{2}} +\cdots~,
\end{eqnarray}
where the $...$ stands for the contributions of the higher states
and continuum. To proceed, besides the transition matrix elements, we need also to know the matrix element $\langle B_Q(p+q,s)\mid
\bar J^{B_Q}(0) \mid
0\rangle$ defined in terms of the residue $\lambda_{B_Q}$,
\begin{eqnarray}\label{matrixel2} \langle B_Q(p+q,s)\mid
\bar J^{B_Q}(0) \mid
0\rangle=\lambda_{B_Q}
\bar u_{B_Q}(p+q,s)~.
\end{eqnarray}
Putting all definitions in Eqs.~(\ref{phys1}) and
(\ref{phys1111}) and using the completeness relation for Dirac particle as
\begin{eqnarray}\label{spinor}
\sum_{s}u_{B_Q}(p+q,s)\overline{u}_{B_Q}(p+q,s)=
\not\!p+\not\!q+m_{B_Q}~,
\end{eqnarray}
we get the following final representations of the correlation functions in physical side:
\begin{eqnarray}\label{sigmaaftera} \Pi_{\mu}^{I}(p,q)\es
\frac{\lambda_{B_Q}u_B(p)}{m_{B_Q}^{2}-(p+q)^{2}}
\Big\{ 2 f_{1}(Q^{2})p_\mu + 2 f_{2}(Q^{2}) p_\mu\not\!q
+\Big[ f_2(Q^2) + f_3(Q^2)\Big] q_\mu\not\!q \nnb \\
\ar 2 g_1(Q^2) p_{\mu}\gamma_5 + 2 g_2(Q^2)p_\mu\not\!q\gamma_5 +
\Big[g_2(Q^2)+g_3(Q^2)\vphantom{\int_0^{x_2}}\Big] q_\mu\not\!q
\gamma_5 \nnb \\
\ar \mbox{\rm other structures}\Big\} +...~, \\ \nnb \\
\label{sigmaafterb} \Pi_{\mu}^{II}(p,q)\es
\frac{\lambda_{B_Q}u_B(p)}{m_{B_Q}^{2}-(p+q)^{2}}
\Big\{ 2 f_{1}^{T}(Q^{2}) p_\mu + 2 f_{2}^{T}(Q^{2})p_\mu\not\!q+
\Big[ f_2^{T}(Q^2) + f_3^{T}(Q^2)\Big]q_\mu\not\!q \nnb \\
\ek 2g_1^{T}(Q^2)p_{\mu}\gamma_5 - 2
g_2^{T}(Q^2)p_\mu\not\!q\gamma_5
-\Big[g_2^{T}(Q^2)+g_3^{T}(Q^2) \Big]q_\mu\not\!q\gamma_5 \nnb \\
\ar \mbox{\rm other structures} \Big\}+... ~,
\end{eqnarray}
where we choose the represented structures to obtain sum rules for the form factors or their combinations. Here, we should comment that besides the presented structures, there are  other structures which one can select
to find the form factors. However, our calculations show that the selected structures lead to the more reliable results having good convergence of sum rules, i.e. in the coefficients of the selected structures,  
contribution of the higher twists is less than those of the lower twists.

Now, we turn our attention to calculate the QCD sides of the
aforesaid correlation functions. They are calculated   in deep
Euclidean region, where $-(p+ q)^2 \rightarrow \infty$. Using the
explicit expressions of the interpolating currents and
contracting out the quark pairs using the Wick's theorem, we find

\begin{eqnarray}\label{mut.m}
\Pi^I_{\mu} &=& \frac{i}{\sqrt{6}} \epsilon^{abc}\int d^4x e^{-iqx}
\Bigg\{\Bigg(\Big[2( C )_{\phi\eta} (\gamma_5)_{\rho\beta}+( C
)_{\phi\beta}
(\gamma_5)_{\rho\eta}+(C)_{\beta\eta}(\gamma_5)_{\rho\phi}\Big]
+\beta\Bigg[2(C \gamma_5 )_{\phi\eta}(I)_{\rho\beta}
 \nonumber \\
&+& (C \gamma_5 )_{\phi\beta}(I)_{\rho\eta}+(C \gamma_5
)_{\beta\eta}(I)_{\rho\phi} \Bigg]\Bigg) \Big[
\gamma_{\mu}(1-\gamma_5)
\Big]_{\sigma\theta}\Bigg\}S_b(-x)_{\beta\sigma}\langle 0 |
s_\eta^a(0) s_\theta^b(x)
u_\phi^c(0) | \Xi (p)\rangle~ ,\nonumber\\
\end{eqnarray}
\begin{eqnarray}\label{mut.mm}
\Pi^{II}_{\mu} &=& \frac{-i}{\sqrt{6}} \epsilon^{abc}\int d^4x
e^{-iqx} \Bigg\{\Bigg(\Big[2( C )_{\phi\eta}
(\gamma_5)_{\rho\beta}+( C )_{\phi\beta}
(\gamma_5)_{\rho\eta}+(C)_{\beta\eta}(\gamma_5)_{\rho\phi}\Big]
+\beta\Bigg[2(C \gamma_5 )_{\phi\eta}(I)_{\rho\beta}
 \nonumber \\
&+& (C \gamma_5 )_{\phi\beta}(I)_{\rho\eta}+(C \gamma_5
)_{\beta\eta}(I)_{\rho\phi} \Bigg]\Bigg) \Big[
i\sigma_{\mu\nu}q^\nu(1-\gamma_5)
\Big]_{\sigma\theta}\Bigg\}S_b(-x)_{\beta\sigma}\langle 0 |
s_\eta^a(0) s_\theta^b(x)
u_\phi^c(0) | \Xi (p)\rangle~ ,\nonumber\\
\end{eqnarray}
for $\Xi_b\rightarrow \Xi l^+l^-$,

\begin{eqnarray}\label{mut.m}
\Pi^I_{\mu} &=& \frac{i}{\sqrt{6}} \epsilon^{abc}\int d^4x e^{-iqx}
\Bigg\{\Bigg(\Big[2( C )_{\phi\eta} (\gamma_5)_{\rho\beta}+( C
)_{\phi\beta}
(\gamma_5)_{\rho\eta}+(C)_{\beta\eta}(\gamma_5)_{\rho\phi}\Big]
+\beta\Bigg[2(C \gamma_5 )_{\phi\eta}(I)_{\rho\beta}
 \nonumber \\
&+& (C \gamma_5 )_{\phi\beta}(I)_{\rho\eta}+(C \gamma_5
)_{\beta\eta}(I)_{\rho\phi} \Bigg]\Bigg) \Big[
\gamma_{\mu}(1-\gamma_5)
\Big]_{\sigma\theta}\Bigg\}S_b(-x)_{\beta\sigma}\langle 0 |
u_\eta^a(0) s_\theta^b(x)
d_\phi^c(0) | \Sigma (p)\rangle~ ,\nonumber\\
\end{eqnarray}
\begin{eqnarray}\label{mut.mm}
\Pi^{II}_{\mu} &=& \frac{-i}{\sqrt{6}} \epsilon^{abc}\int d^4x
e^{-iqx} \Bigg\{\Bigg(\Big[2( C )_{\phi\eta}
(\gamma_5)_{\rho\beta}+( C )_{\phi\beta}
(\gamma_5)_{\rho\eta}+(C)_{\beta\eta}(\gamma_5)_{\rho\phi}\Big]
+\beta\Bigg[2(C \gamma_5 )_{\phi\eta}(I)_{\rho\beta}
 \nonumber \\
&+& (C \gamma_5 )_{\phi\beta}(I)_{\rho\eta}+(C \gamma_5
)_{\beta\eta}(I)_{\rho\phi} \Bigg]\Bigg) \Big[
i\sigma_{\mu\nu}q^\nu(1-\gamma_5)
\Big]_{\sigma\theta}\Bigg\}S_b(-x)_{\beta\sigma}\langle 0 |
u_\eta^a(0) s_\theta^b(x)
d_\phi^c(0) | \Sigma (p)\rangle~ ,\nonumber\\
\end{eqnarray}
for $\Xi_b\rightarrow \Sigma l^+l^-$,

\begin{eqnarray}
\Pi^I_{\mu} &=& \frac{i}{\sqrt{6}} \epsilon^{abc}\int d^4x e^{-iqx}
\Bigg\{\Bigg(\Big[2( C )_{\phi\eta} (\gamma_5)_{\rho\beta}+( C
)_{\phi\beta}
(\gamma_5)_{\rho\eta}+(C)_{\beta\eta}(\gamma_5)_{\rho\phi}\Big]
+\beta\Bigg[2(C \gamma_5 )_{\phi\eta}(I)_{\rho\beta}
 \nonumber \\
&+& (C \gamma_5 )_{\phi\beta}(I)_{\rho\eta}+(C \gamma_5
)_{\beta\eta}(I)_{\rho\phi} \Bigg]\Bigg) \Big[
\gamma_{\mu}(1-\gamma_5)
\Big]_{\sigma\theta}\Bigg\}S_b(-x)_{\beta\sigma}\langle 0 |
u_\eta^a(0) s_\theta^b(x)
d_\phi^c(0) | \Sigma (p)\rangle~ ,\nonumber\\
\end{eqnarray}
\begin{eqnarray}
\Pi^{II}_{\mu} &=& \frac{-i}{\sqrt{6}} \epsilon^{abc}\int d^4x
e^{-iqx} \Bigg\{\Bigg(\Big[2( C )_{\phi\eta}
(\gamma_5)_{\rho\beta}+( C )_{\phi\beta}
(\gamma_5)_{\rho\eta}+(C)_{\beta\eta}(\gamma_5)_{\rho\phi}\Big]
+\beta\Bigg[2(C \gamma_5 )_{\phi\eta}(I)_{\rho\beta}
 \nonumber \\
&+& (C \gamma_5 )_{\phi\beta}(I)_{\rho\eta}+(C \gamma_5
)_{\beta\eta}(I)_{\rho\phi} \Bigg]\Bigg) \Big[
i\sigma_{\mu\nu}q^\nu(1-\gamma_5)
\Big]_{\sigma\theta}\Bigg\}S_b(-x)_{\beta\sigma}\langle 0 |
u_\eta^a(0) s_\theta^b(x)
d_\phi^c(0) | \Sigma (p)\rangle~ ,\nonumber\\
\end{eqnarray}
for $\Xi_c\rightarrow \Sigma l^+l^-$,

\begin{eqnarray}\label{mut.m}
\Pi^I_{\mu} &=& \frac{-i}{\sqrt{2}} \epsilon^{abc}\int d^4x e^{-iqx}
\Bigg\{\Bigg(\Big[( C )_{\phi\beta} (\gamma_5)_{\rho\eta}-( C
)_{\beta\eta} (\gamma_5)_{\rho\phi}\Big] +\beta\Bigg[(C \gamma_5
)_{\phi\beta}(I)_{\rho\eta}
 \nonumber \\
&-& (C \gamma_5 )_{\beta\eta}(I)_{\rho\phi}\Bigg]\Bigg) \Big[
\gamma_{\mu}(1-\gamma_5)
\Big]_{\sigma\theta}\Bigg\}S_b(-x)_{\beta\sigma}\langle 0 |
s_\eta^a(0) s_\theta^b(x)
u_\phi^c(0) | \Xi (p)\rangle~ ,\nonumber\\
\end{eqnarray}
\begin{eqnarray}\label{mut.mm}
\Pi^{II}_{\mu} &=& \frac{i}{\sqrt{2}} \epsilon^{abc}\int d^4x
e^{-iqx} \Bigg\{\Bigg(\Big[( C )_{\phi\beta} (\gamma_5)_{\rho\eta}-(
C )_{\beta\eta} (\gamma_5)_{\rho\phi}\Big] +\beta\Bigg[(C \gamma_5
)_{\phi\beta}(I)_{\rho\eta}
 \nonumber \\
&-& (C \gamma_5 )_{\beta\eta}(I)_{\rho\phi}\Bigg]\Bigg) \Big[
i\sigma_{\mu\nu}q^\nu(1-\gamma_5)
\Big]_{\sigma\theta}\Bigg\}S_b(-x)_{\beta\sigma}\langle 0 |
s_\eta^a(0) s_\theta^b(x)
u_\phi^c(0) | \Xi (p)\rangle~ ,\nonumber\\
\end{eqnarray}
for $\Xi'_b\rightarrow \Xi l^+l^-$,

\begin{eqnarray}\label{mut.m}
\Pi^I_{\mu} &=& \frac{-i}{\sqrt{2}} \epsilon^{abc}\int d^4x e^{-iqx}
\Bigg\{\Bigg(\Big[( C )_{\phi\beta} (\gamma_5)_{\rho\eta}-( C
)_{\beta\eta} (\gamma_5)_{\rho\phi}\Big] +\beta\Bigg[(C \gamma_5
)_{\phi\beta}(I)_{\rho\eta}
 \nonumber \\
&-& (C \gamma_5 )_{\beta\eta}(I)_{\rho\phi}\Bigg]\Bigg) \Big[
\gamma_{\mu}(1-\gamma_5)
\Big]_{\sigma\theta}\Bigg\}S_b(-x)_{\beta\sigma}\langle 0 |
d_\eta^a(0) s_\theta^b(x)
d_\phi^c(0) | \Sigma (p)\rangle~ ,\nonumber\\
\end{eqnarray}
\begin{eqnarray}\label{mut.mm}
\Pi^{II}_{\mu} &=& \frac{i}{\sqrt{2}} \epsilon^{abc}\int d^4x
e^{-iqx} \Bigg\{\Bigg(\Big[( C )_{\phi\beta} (\gamma_5)_{\rho\eta}-(
C )_{\beta\eta} (\gamma_5)_{\rho\phi}\Big] +\beta\Bigg[(C \gamma_5
)_{\phi\beta}(I)_{\rho\eta}
 \nonumber \\
&-& (C \gamma_5 )_{\beta\eta}(I)_{\rho\phi}\Bigg]\Bigg) \Big[
i\sigma_{\mu\nu}q^\nu(1-\gamma_5)
\Big]_{\sigma\theta}\Bigg\}S_b(-x)_{\beta\sigma}\langle 0 |
d_\eta^a(0) s_\theta^b(x)
d_\phi^c(0) | \Sigma (p)\rangle~ ,\nonumber\\
\end{eqnarray}
for $\Xi'_{b}\rightarrow \Sigma l^+l^-$,

\begin{eqnarray}\label{mut.m}
\Pi^I_{\mu} &=& \frac{-i}{\sqrt{2}} \epsilon^{abc}\int d^4x e^{-iqx}
\Bigg\{\Bigg(\Big[( C )_{\phi\beta} (\gamma_5)_{\rho\eta}-( C
)_{\beta\eta} (\gamma_5)_{\rho\phi}\Big] +\beta\Bigg[(C \gamma_5
)_{\phi\beta}(I)_{\rho\eta}
 \nonumber \\
&-& (C \gamma_5 )_{\beta\eta}(I)_{\rho\phi}\Bigg]\Bigg) \Big[
\gamma_{\mu}(1-\gamma_5)
\Big]_{\sigma\theta}\Bigg\}S_c(-x)_{\beta\sigma}\langle 0 |
u_\eta^a(0) s_\theta^b(x)
d_\phi^c(0) | \Sigma (p)\rangle~ ,\nonumber\\
\end{eqnarray}
\begin{eqnarray}\label{mut.mm}
\Pi^{II}_{\mu} &=& \frac{i}{\sqrt{2}} \epsilon^{abc}\int d^4x
e^{-iqx} \Bigg\{\Bigg(\Big[( C )_{\phi\beta} (\gamma_5)_{\rho\eta}-(
C )_{\beta\eta} (\gamma_5)_{\rho\phi}\Big] +\beta\Bigg[(C \gamma_5
)_{\phi\beta}(I)_{\rho\eta}
 \nonumber \\
&-& (C \gamma_5 )_{\beta\eta}(I)_{\rho\phi}\Bigg]\Bigg) \Big[
i\sigma_{\mu\nu}q^\nu(1-\gamma_5)
\Big]_{\sigma\theta}\Bigg\}S_c(-x)_{\beta\sigma}\langle 0 |
u_\eta^a(0) s_\theta^b(x)
d_\phi^c(0) | \Sigma (p)\rangle~ ,\nonumber\\
\end{eqnarray}
for $\Xi'_{c}\rightarrow \Sigma l^+l^-$,

\begin{eqnarray}\label{Eq11}
\Pi_\mu &=& \frac{i}{\sqrt{6}} \epsilon^{abc}\int d^4x e^{-iqx}
\Bigg\{\Big[2 ( C )_{\phi\eta} (\gamma_5)_{\rho\beta}+( C
)_{\phi\beta} (\gamma_5)_{\rho\eta}+( C )_{\beta\eta}
(\gamma_5)_{\rho\phi}\Big] +\beta\Bigg[2 (C \gamma_5
)_{\phi\eta}(I)_{\rho\beta}
 \nonumber \\
&+& (C \gamma_5 )_{\phi\beta}(I)_{\rho\eta}+(C \gamma_5
)_{\beta\eta}(I)_{\rho\phi} \Bigg]\Bigg\} \Big[ \gamma_{\mu}
(1-\gamma_5) \Big]_{\sigma\theta}S_c(-x)_{\beta\sigma} \langle  0 |
s_\eta^a(0)
 s_\theta^b(x)   d_\phi^c(0) | \Xi(p)\rangle ,\nonumber\\
\end{eqnarray}
for $\Xi_c\rightarrow \Xi l\nu$,

\begin{eqnarray}\label{Eq12}
\Pi_\mu &=& \frac{i}{\sqrt{6}} \epsilon^{abc}\int d^4x e^{-iqx}
\Bigg\{\Big[2 ( C )_{\phi\eta} (\gamma_5)_{\rho\beta}+( C
)_{\phi\beta} (\gamma_5)_{\rho\eta}+( C )_{\beta\eta}
(\gamma_5)_{\rho\phi}\Big] +\beta\Bigg[2 (C \gamma_5
)_{\phi\eta}(I)_{\rho\beta}
 \nonumber \\
&+& (C \gamma_5 )_{\phi\beta}(I)_{\rho\eta}+(C \gamma_5
)_{\beta\eta}(I)_{\rho\phi} \Bigg]\Bigg\} \Big[
\gamma_{\mu}(1-\gamma_5) \Big]_{\sigma\theta}S_c(-x)_{\beta\sigma}
\langle  0 |  u_\eta^a(0)
 s_\theta^b(x)   d_\phi^c(0) | \Sigma(p)\rangle ,\nonumber\\
\end{eqnarray}
for $\Xi_c\rightarrow \Sigma l\nu$,

\begin{eqnarray}\label{Eq11}
\Pi_\mu &=& \frac{-i}{\sqrt{2}} \epsilon^{abc}\int d^4x e^{-iqx}
\Bigg\{\Big[ ( C )_{\phi\beta} (\gamma_5)_{\rho\eta}-( C
)_{\beta\eta} (\gamma_5)_{\rho\phi}\Big] +\beta\Bigg[ (C \gamma_5
)_{\phi\beta}(I)_{\rho\eta}
 - (C \gamma_5 )_{\beta\eta}(I)_{\rho\phi}\Bigg]\Bigg\}
\nonumber \\
&&  \Big[ \gamma_{\mu}(1-\gamma_5)
\Big]_{\sigma\theta}S_c(-x)_{\beta\sigma} \langle  0 |  s_\eta^a(0)
 s_\theta^b(x)   u_\phi^c(0) | \Xi(p)\rangle ,\nonumber\\
\end{eqnarray}
for $\Xi'_c\rightarrow \Xi l\nu$, and

\begin{eqnarray}\label{Eq11}
\Pi_\mu &=& \frac{-i}{\sqrt{2}} \epsilon^{abc}\int d^4x e^{-iqx}
\Bigg\{\Big[ ( C )_{\phi\beta} (\gamma_5)_{\rho\eta}-( C
)_{\beta\eta} (\gamma_5)_{\rho\phi}\Big] +\beta\Bigg[ (C \gamma_5
)_{\phi\beta}(I)_{\rho\eta}
 - (C \gamma_5 )_{\beta\eta}(I)_{\rho\phi}\Bigg]\Bigg\}
\nonumber \\
&&  \Big[ \gamma_{\mu}(1-\gamma_5)
\Big]_{\sigma\theta}S_c(-x)_{\beta\sigma} \langle  0 |  u_\eta^a(0)
 s_\theta^b(x)   d_\phi^c(0) | \Sigma(p)\rangle ,\nonumber\\
\end{eqnarray}
for $\Xi'_c\rightarrow \Sigma l\nu$, where $ S_Q(x)$ is the heavy
quark propagator which is given by \cite{22Balitsky}:
\begin{eqnarray}\label{heavylightguy}
 S_Q (x)& =&  S_Q^{free} (x) - i g_s \int \frac{d^4 k}{(2\pi)^4}
e^{-ikx} \int_0^1 dv \Bigg[\frac{\not\!k + m_Q}{( m_Q^2-k^2)^2}
G^{\mu\nu}(vx) \sigma_{\mu\nu} \nnb \\
\ar \frac{1}{m_Q^2-k^2} v x_\mu G^{\mu\nu} \gamma_\nu \Bigg]~,
\end{eqnarray}
and,
\begin{eqnarray}\label{freeprop} S^{free}_{Q}
\es\frac{m_{Q}^{2}}{4\pi^{2}}\frac{K_{1}(m_{b}\sqrt{-x^2})}{\sqrt{-x^2}}-i
\frac{m_{Q}^{2}\not\!x}{4\pi^{2}x^2}K_{2}(m_{b}\sqrt{-x^2})~,
\end{eqnarray}
 with
$K_i$ being the Bessel functions. In
Eq. (\ref{heavylightguy}), the $S_Q^{free}$ corresponds to the free
propagation of the heavy quark. The interaction of the heavy quark
with the external gluon field is represented by the remaining terms.
However  calculation of these types of interactions requires 
knowledge of the currently unknown four-and five-particle baryonic
DA's. The  contribution of such terms are
expected to be small \cite{eky1,eky2,eky3}, hence, in the present work we ignore their contributions.

To complete the calculations in
QCD side, we need also the wave functions of the $\Xi$ and $\Sigma$
baryons, i.e., $\epsilon^{abc} \langle 0 | s_\eta^a(0) s_\theta^b(x)
u_\phi^c(0) | \Xi (p)\rangle$ and $\epsilon^{abc} \langle 0 |
u(d)_\eta^a(0) s_\theta^b(x) d_\phi^c(0) | \Sigma (p)\rangle$. These
wave functions are expanded in terms of DA's having different twists
which  are calculated in \cite{Yong-LuLiu} and \cite{Liu}.  For
completeness, we present the explicit forms of the wave
functions together with the DA's in the Appendix. Using the wave
functions and heavy quark propagator we obtain the correlation
functions in QCD side. 

To obtain sum rules for the  form factors, we match the coefficients
of the same Dirac structures from both sides of the
correlation functions. We also apply Borel transformation and
continuum subtraction to suppress the contribution
 of the higher states and continuum. These processes bring us two auxiliary parameters, namely Borel mass parameter $M^2$ and continuum threshold $s_0$ which we will find the working regions for these quantities in the next section. In the meanwhile, we need also
the residues $\lambda_{\Xi_Q(\Xi'_Q)}$ whose explicit forms are
given in \cite{T. M. Aliev}. The explicit forms of sum
rules for the form factors are very lengthy and we do not present
their explicit expressions here, but we will give their
fit functions in terms of $q^2$ in next section.

\section{Numerical Results}
In this section, we numerically analyze the form factors and obtain
their behavior in terms of $q^2$. Using the fit functions of the
form factors, we also calculate the decay rates for all considered
channels and branching ratios for the channels in which
the lifetime of initial particle  is known. Some input parameters
used in the numerical calculations are: $m_{\Xi_{b}^0} = (5790.5
\pm2.7)$~MeV, $m_{\Xi^{'}_{b}} = (5790.5 \pm2.7)$~MeV, $m_{\Xi^0} =
(1314.86 \pm0.20)$~MeV, $m_{\Xi_{c}^0} =
(2470.88^{+0.34}_{-0.80})$~MeV, $m_{\Xi^{'^+}_{c}} = (2575.6
\pm3.1)$~MeV, $m_{\Xi^{'^0}_{c}} = (2577.9 \pm2.9)$~MeV,
$m_{\Sigma^{0}} = (1192.642 \pm0.024)$~MeV, $m_{\Sigma^{-}} =
(1197.449 \pm0.030)$~MeV, $m_b = (4.7\pm 0.1)$~GeV, $m_c =
(1.27^{+0.07}_{-0.09})$~GeV, $|V_{cs} |=1.023\pm0.036$, $|V_{cd}
|=0.230\pm0.011$, $|V_{tb}V_{td*} |=8.27\times 10^{-3}$,
$|V_{tb}V_{ts*} |=0.041$ , $V_{bc}=(41.2\pm1.1)\times10^{-3}$,
$V_{bu}=(3.93\pm0.36)\times10^{-3}$ \cite{Rstp01}, $C_{7}^{eff}=-0.313$, $C_{9}^{eff}=4.344$ and $C_{10}=-4.669$ \cite{bura}.

The main input parameters of the LCSR for form factors are  the DA's
of the  $\Xi$ and $\Sigma$ baryons presented in the Appendix. These
DA's contain also four independent parameters.  These parameters in
the case of $\Xi$ baryon are given as \cite{Yong-LuLiu}:
\begin{eqnarray}
f_{\Xi}&=&(9.9\pm0.4)\times10^{-3}\; \mbox{GeV}^2,
\hspace{2.5cm}\lambda_1=-(2.8\pm0.1)\times10^{-2}\; \mbox{GeV}^2,\nonumber\\
\lambda_2&=&(5.2\pm0.2)\times10^{-2}\;
\mbox{GeV}^2,\hspace{2.5cm}\lambda_3=(1.7\pm0.1)\times10^{-2}\;
\mbox{GeV}^2,\label{Xipara}
\end{eqnarray}
and for $\Sigma$ baryon, they take the values \cite{Liu}:
\begin{eqnarray}
f_{\Sigma}&=&(9.4\pm0.4)\times10^{-3}\; \mbox{GeV}^2,\hspace{2.5cm}
\lambda_1=-(2.5\pm0.1)\times10^{-2}\; \mbox{GeV}^2,\nonumber\\
\lambda_2&=&(4.4\pm0.1)\times10^{-2}\;
\mbox{GeV}^2,\hspace{2.5cm}\lambda_3=(2.0\pm0.1)\times10^{-2}\;
\mbox{GeV}^2.\label{sigmapara}
\end{eqnarray}

\begin{figure}[h!]
\begin{center}
\includegraphics[width=10cm]{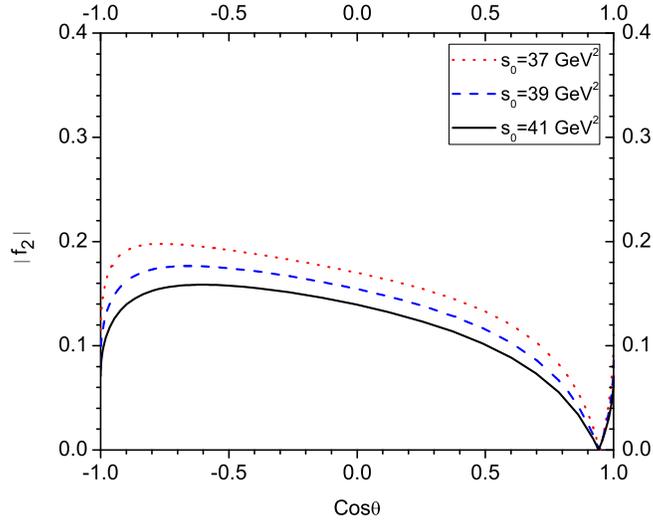}
\end{center}
\caption{Form factor $f_2$ as a function of  $\cos\theta$ for
$\Xi_b\rightarrow\Xi \ell^+ \ell^-$ decay at $q^2=13 ~GeV^2$ and    working region  of $M^2$ . }
\label{f2CosTheta}
\end{figure}

\begin{figure}[h!]
\begin{center}
\includegraphics[width=10cm]{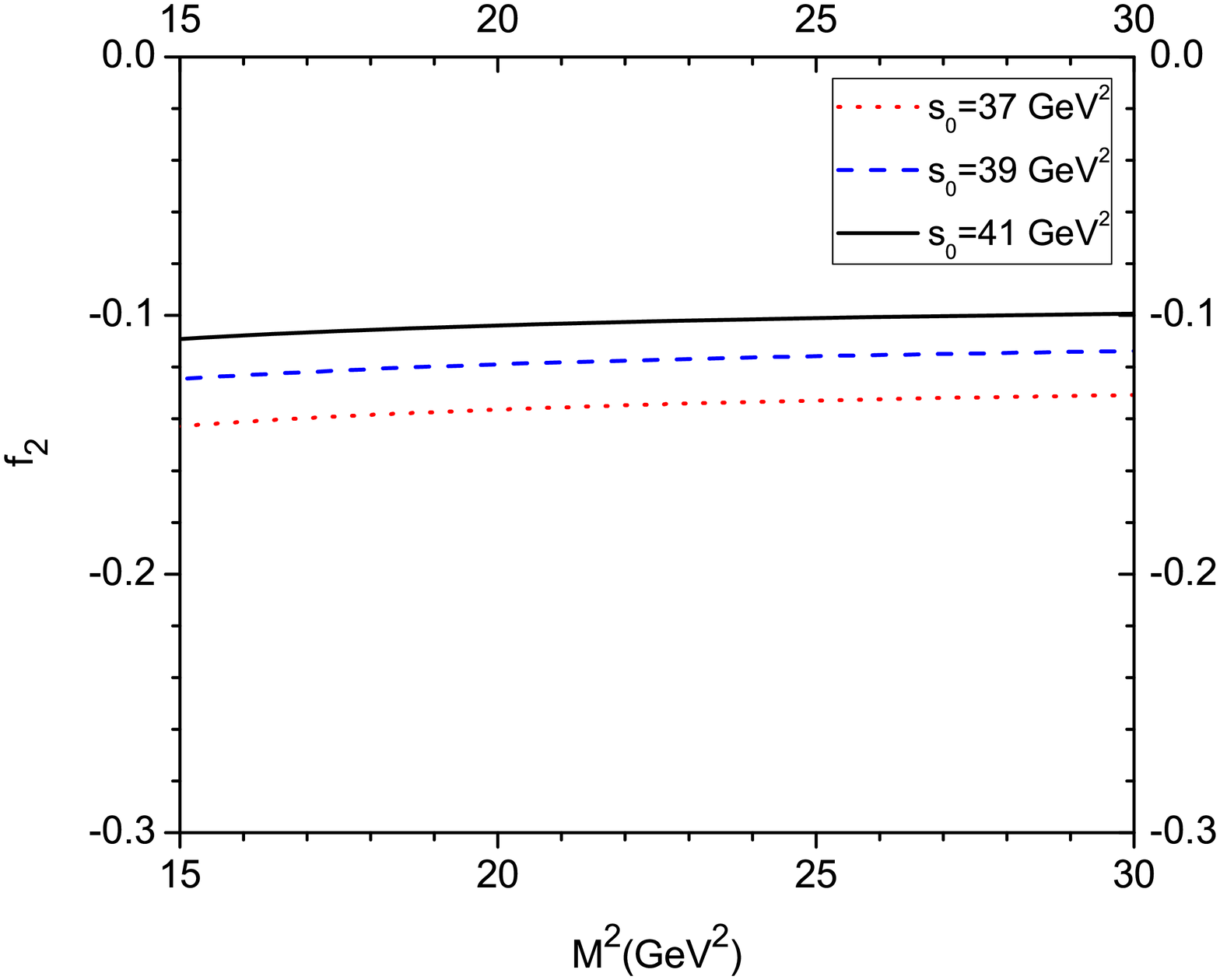}
\end{center}
\caption{Form factor $f_2$ as a function of  Borel mass parameter $M^2$ for
$\Xi_b\rightarrow\Xi l^+ l^-$ decay at $q^2=13~ GeV^2$ and  working region  of $\beta$. }
\label{f2MB}
\end{figure}

The LCSR for form factors contain also three auxiliary parameters.
Borel mass parameter $M^2$ and continuum threshold $s_0$ are two of
them coming from the Borel transformation and continuum subtraction,
respectively. The general parameter $\beta$ is the third parameter
entering  the calculations from the general form of the
interpolating currents for $B_Q$ baryons. According to the standard
criteria in QCD sum rules,
 the results of form factors should be independent of these auxiliary parameters. Hence, we should look for working regions of these parameters such that  the dependence of the results on these parameters are weak.
 The working
region for the Borel mass parameter is
 determined requiring that not only the higher states and continuum
contributions constitute  a small percentage of the total dispersion
integral but also the series of the light cone expansion with
increasing twist should  converge. This leads to the
interval $15~\mbox{GeV}^2\leq M^2\leq30~\mbox{GeV}^2$
for bottom baryons and  $4~\mbox{GeV}^2\leq M^2\leq10~\mbox{GeV}^2$
for charmed baryons. The continuum threshold $s_0$ is not totally
arbitrary but it is related to the energy of the first excited
state. Our numerical calculations show that in the region
$(m_{B_Q}+0.3)^2~\mbox{GeV}^2\leq
s_0\leq(m_{B_Q}+0.7)^2~\mbox{GeV}^2$, the results of the form
factors exhibit very weak dependency on this parameter. Our
numerical calculations also lead to the  working region
 $-0.6\leq \cos\theta\leq0.3$ with $\tan \theta=\beta$  for the general parameter $\beta$. 
As an example,   we present the dependence of the form factor  $f_2$
for $\Xi_b\rightarrow\Xi \ell^+ \ell^-$ on $\cos\theta$ and $M^2$ in
Figures \ref{f2CosTheta} and \ref{f2MB}, respectively. From these
figures, we see that the form factor $f_2$ depends weakly on the $M^2$ and $s_0$ compared to the $\cos\theta$. However, the dependence of the $f_2$ on $\cos\theta$ in the above mentioned working 
region is minimal compared to the intervals out of the working region.

Now, we proceed to find the $q^2$ dependence of the form factors in whole physical region, i.e. $4m_\ell^2 \leq q^2\leq (m_{B_Q}- m_B)^2$ for loop level and
 $m_\ell^2 \leq q^2\leq (m_{B_Q}- m_B)^2$ for tree level transitions. However, unfortunately the sum rules for form factors are truncated at
some points  and are not reliable in the whole physical region. This point  for instance for the $\Xi_b\rightarrow\Xi l^+ l^-$ transition is
roughly at $q^2=15~ GeV^2$.  To extend the results to whole physical region, we look for 
parametrization of the form factors such that in the reliable region, the results obtained from fit parametrization coincide with the sum rules predictions. Using the above working regions for the auxiliary
parameters as well as other input parameters, we find that the form
factors are well extrapolated by
 the fit parametrization,
\begin{eqnarray}
f_i(q^2)[g_i(q^2)]=\frac{a}{(1-\frac{q^2}{m_{fit}^2})}+\frac{b}{(1-\frac{q^2}{m_{fit}^2})^2}
\label{parametrization1}.
\end{eqnarray}
The central values for the fit parameters $a$, $b$, and $m_{fit}$ as well as values of the form factors at $q^2=0$ are presented in
Tables~\ref{tab:13}-\ref{tab:181}. The errors in the values of the
form factors at $q^2=0$ are due to the variation of the auxiliary
parameters $M^2$, $s_0$, and $\beta$ in their working regions as
well as the errors in the other input parameters. To see how  the results obtained from the fit function coincide well with the sum rules predictions at reliable region, we
depict the dependence of the form factors $f_2$ and $f_2^T$, as examples, on $q^2$ in figures 4 and 5. From these figures, we see that the results obtained from the fit parametrization 
describe well the sum rules results in the reliable region.
\begin{figure}[h!]
\begin{center}
\includegraphics[width=10cm]{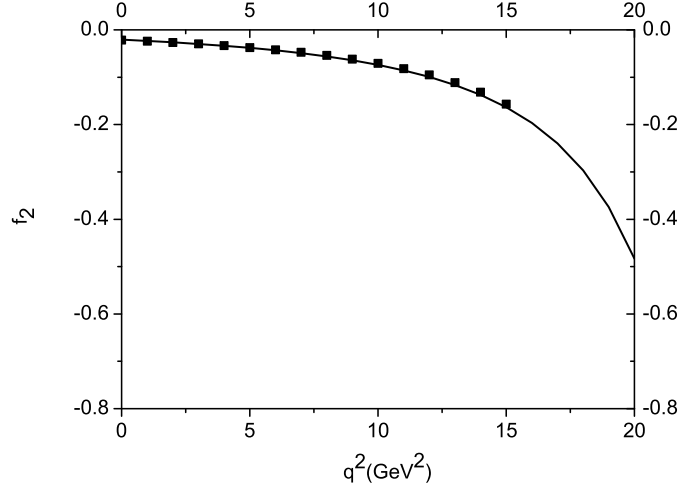}
\end{center}
\caption{Form factor $f_2$ as a function of   $q^2$ for
$\Xi_b\rightarrow\Xi l^+ l^-$ decay at  working regions  of auxiliary parameters. The boxes show the sum rules predictions and the solid line belongs to the result obtained from 
fit parametrization.}
\label{}
\end{figure}
\begin{figure}[h!]
\begin{center}
\includegraphics[width=10cm]{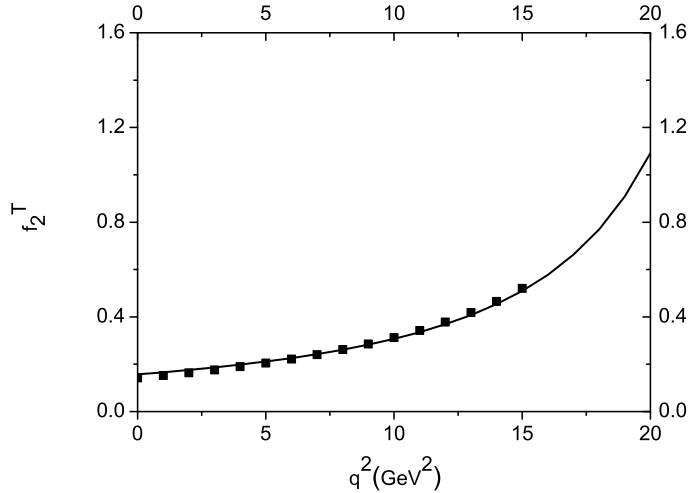}
\end{center}
\caption{Form factor $f^T_2$ as a function of   $q^2$ for
$\Xi_b\rightarrow\Xi l^+ l^-$ decay at  working regions  of auxiliary parameters. The boxes show the sum rules predictions and the solid line belongs to the result obtained from 
fit parametrization.}
\label{}
\end{figure}

Our last task is to calculate the decay rates and branching ratios of the considered channels using the fit functions of the form factors. Considering the amplitudes of the transitions and definitions of
the transition matrix elements in terms of form factors, the differential decay rate for loop level transitions is obtained as \cite{Azizi3} :
\begin{eqnarray} \frac{d\Gamma}{ds} = \frac{G_F^2\alpha^2_{em}
m_{B_Q}}{8192 \pi^5}| V_{Q'Q}V_{Q'q}^*|^2 v
\sqrt{\lambda} \, \Bigg[ \Theta(s) + \frac{1}{3} \Delta(s)\Bigg]~,
\label{rate} \end{eqnarray} where $s=q^2/m^2_{B_Q}$,
 $G_F = 1.17 \times 10^{-5}$ GeV$^{-2}$,  $\lambda=\lambda(1, r, s)$ with $
\lambda(a,b,c)=a^2+b^2+c^2-2ab-2ac-2bc$ and $v=\sqrt{1-\frac{4 m_\ell^2}{q^2}}$ is the
lepton velocity.  The functions $\Theta(s)$ and $\Delta(s)$ are
given as:

\bea \Theta(s) \es 32 m_\ell^2 m_{B_Q}^4 s (1+r-s)
\ga \vel D_3 \ver^2 +
\vel E_3 \ver^2 \dr \nnb \\
\ar 64 m_\ell^2 m_{B_Q}^3 (1-r-s) \, \mbox{\rm Re}
[D_1^\ast E_3 + D_3
E_1^\ast] \nnb \\
\ar 64 m_{B_Q}^2 \sqrt{r} (6 m_\ell^2 -
m_{B_Q}^2 s)
{\rm Re} [D_1^\ast E_1] \nnb \\
\ar 64 m_\ell^2 m_{B_Q}^3 \sqrt{r} \Big( 2
m_{B_Q} s {\rm Re} [D_3^\ast E_3] + (1 - r + s)
{\rm Re} [D_1^\ast D_3 + E_1^\ast E_3]\Big) \nnb \\
\ar 32 m_{B_Q}^2 (2 m_\ell^2 +
m_{B_Q}^2 s) \Big\{ (1 - r + s)
m_{B_Q} \sqrt{r} \,
\mbox{\rm Re} [A_1^\ast A_2 + B_1^\ast B_2] \nnb \\
\ek m_{B_Q} (1 - r - s) \, \mbox{\rm Re} [A_1^\ast
B_2 + A_2^\ast B_1] - 2 \sqrt{r} \Big( \mbox{\rm Re} [A_1^\ast B_1]
+ m_{B_Q}^2 s \,
\mbox{\rm Re} [A_2^\ast B_2] \Big) \Big\} \nnb \\
\ar 8 m_{B_Q}^2 \Big\{ 4 m_\ell^2 (1 + r - s) +
m_{B_Q}^2 \Big[(1-r)^2 - s^2 \Big]
\Big\} \ga \vel A_1 \ver^2 +  \vel B_1 \ver^2 \dr \nnb \\
\ar 8 m_{B_Q}^4 \Big\{ 4 m_\ell^2 \Big[ \lambda + (1
+ r - s) s \Big] + m_{B_Q}^2 s \Big[(1-r)^2 - s^2
\Big]
\Big\} \ga \vel A_2 \ver^2 +  \vel B_2 \ver^2 \dr \nnb \\
\ek 8 m_{B_Q}^2 \Big\{ 4 m_\ell^2 (1 + r - s) -
m_{B_Q}^2 \Big[(1-r)^2 - s^2 \Big]
\Big\} \ga \vel D_1 \ver^2 +  \vel E_1 \ver^2 \dr \nnb \\
\ar 8 m_{B_Q}^5 s v^2 \Big\{ - 8
m_{B_Q} s \sqrt{r}\, \mbox{\rm Re} [D_2^\ast E_2] +
4 (1 - r + s) \sqrt{r} \, \mbox{\rm Re}[D_1^\ast D_2+E_1^\ast E_2]\nnb \\
\ek 4 (1 - r - s) \, \mbox{\rm Re}[D_1^\ast E_2+D_2^\ast E_1] +
m_{B_Q} \Big[(1-r)^2 -s^2\Big] \ga \vel D_2 \ver^2 +
\vel E_2 \ver^2\dr \Big\}~,
\eea \bea \Delta \left( s\right) \es - 8 m_{B_Q}^4
v^2 \lambda \ga \vel A_1 \ver^2 + \vel B_1 \ver^2 + \vel D_1 \ver^2
+ \vel E_1 \ver^2 \dr \nnb \\
\ar 8 m_{B_Q}^6 s v^2 \lambda \Big( \vel A_2 \ver^2 +
\vel B_2 \ver^2 + \vel D_2 \ver^2 + \vel E_2 \ver^2  \Big)~, \eea
where $r= m^2_{B}/m^2_{B_Q}$  and
 \bea \label{a9} A_1 \es
\frac{1}{q^2}\ga
f_1^T+g_1^T \dr \ga -2 m_{Q} C_7\dr + \ga f_1-g_1 \dr C_9^{eff} \nnb \\
A_2 \es A_1 \ga 1 \rar 2 \dr ~,\nnb \\
A_3 \es A_1 \ga 1 \rar 3 \dr ~,\nnb \\
B_1 \es A_1 \ga g_1 \rar - g_1;~g_1^T \rar - g_1^T \dr ~,\nnb \\
B_2 \es B_1 \ga 1 \rar 2 \dr ~,\nnb \\
B_3 \es B_1 \ga 1 \rar 3 \dr ~,\nnb \\
D_1 \es \ga f_1-g_1 \dr C_{10} ~,\nnb \\
D_2 \es D_1 \ga 1 \rar 2 \dr ~,\nnb \\
D_3 \es D_1 \ga 1 \rar 3 \dr ~,\nnb \\
E_1 \es D_1 \ga g_1 \rar - g_1 \dr ~,\nnb \\
E_2 \es E_1 \ga 1 \rar 2 \dr ~,\nnb \\
E_3 \es E_1 \ga 1 \rar 3 \dr ~. \eea
Integrating the differential decay rate over $s$ in whole physical region, $ 4m_\ell^2/m^2_{B_Q} \leq s\leq (1- \sqrt{r})^2$, one can obtain the total decay rate.

For the tree level transitions, the formula for the decay width is given by \cite{Faessler,Pietschmann:1974ap}:
\bea\label{Gamma_BiBf} \Gamma  = \frac{G_F^2}{384 \pi^3
m_{\Xi^{(')}_{c}}^3} \ |V_{\rm cs(d)}|^2 \, \,
\int\limits_{m_l^2}^{\delta^2} dq^2  \ (1 - m_l^2/q^2)^2 \
\sqrt{(\sigma^2 - q^2) (\delta^2 - q^2)} \ N(q^2) \eea
where
\bea N(q^2) &=& F_1^2(q^2) (\delta^2 (4q^2 - m_l^2) + 2 \sigma^2
\delta^2
(1 + 2 m_l^2/q^2) - (\sigma^2 + 2q^2) (2q^2 + m_l^2) )\nonumber\\[3mm]
&+& F_2^2(q^2) (\delta^2 - q^2)(2 \sigma^2 + q^2) (2q^2 +
m_l^2)/m_{\Xi^{(')}_{c}}^2
+ 3 F_3^2(q^2) m_l^2 (\sigma^2 - q^2) q^2/m_{\Xi^{(')}_{c}}^2  \nonumber\\[3mm]
&+& 6 F_1(q^2) F_2(q^2) (\delta^2 - q^2) (2 q^2 + m_l^2)
\sigma/m_{\Xi^{(')}_{c}}
- 6 F_1(q^2) F_3(q^2)  m_l^2 (\sigma^2 - q^2) \delta/m_{\Xi^{(')}_{c}} \nonumber\\[3mm]
&+& G_1^2(q^2) (\sigma^2 (4q^2 - m_l^2) + 2 \sigma^2 \delta^2
(1 + 2 m_l^2/q^2) - (\delta^2 + 2q^2) (2q^2 + m_l^2) )\nonumber\\[3mm]
&+& G_2^2(q^2) (\sigma^2 - q^2)(2 \delta^2 + q^2) (2q^2 +
m_l^2)/m_{\Xi^{(')}_{c}}^2
+ 3 G_3^2(q^2) m_l^2 (\delta^2 - q^2) q^2/m_{\Xi^{(')}_{c}}^2  \nonumber\\[3mm]
&-& 6 G_1(q^2) G_2(q^2) (\sigma^2 - q^2) (2 q^2 + m_l^2)
\delta/m_{\Xi^{(')}_{c}}
 +  6 G_1(q^2) G_3(q^2)  m_l^2 (\delta^2 - q^2) \sigma/m_{\Xi^{(')}_{c}}  \,,\nonumber\\
\eea
 with $F_1(q^2)=f_1(q^2)$, $F_2(q^2)=m_{\Xi^{(')}_{c}}f_2(q^2)$, $F_3(q^2)=m_{\Xi^{(')}_{c}}f_3(q^2)$,
 $G_1(q^2)=g_1(q^2)$, $G_2(q^2)=m_{\Xi^{(')}_{c}}g_2(q^2)$,
 $G_3(q^2)=m_{\Xi^{(')}_{c}}g_3(q^2)$, $\sigma  =
m_{\Xi^{(')}_{c}} + m_{B}$, $\delta  = m_{\Xi^{(')}_{c}} - m_{B}$
and $m_l$ is the lepton's mass. The numerical results of  decay width
for considered channels are presented in Table~\ref{tab:27}.
Finally, for the channels which we know the lifetime of the initial
particles \cite{Rstp01}, we calculate the branching ratios as
presented in  Table~\ref{tab:27777}. The orders of branching
fractions for most of the channels presented in
 Table~\ref{tab:27777} show that these channels are accessible at LHC.

\section{Conclusion}
In the present study, we have considered various  loop level and tree level semileptonic decays of heavy   $\Xi'_{b(c)}$ and   $\Xi_{b(c)}$  baryons
 to the light  $\Xi$ and $\Sigma $ baryons
 in the framework of the light cone QCD sum rules. The most general form of the interpolating currents for the considered heavy baryons as well as the recently available
 distribution amplitudes of the
 $\Xi$ and $\Sigma $ baryons have been used to calculate twelve  form factors for loop level and six form factors for tree level transitions in full theory of QCD.
 Using the sum rules for the form factors, then, we have evaluated 
 the decay rates of the related transitions. For those transitions with known lifetime, we have also calculated their branching fractions. The orders of branching fractions for
tree level $\Xi_{c} \rar \Sigma ~l^+\nu_{l}$ and $\Xi_{c} \rar \Xi~ l^+\nu_{l}$ (with $l=e$ or $\mu$) as well as rare loop level $\Xi_{b} \rar \Xi~ l^+l^-$ 
and $\Xi_{b} \rar \Sigma~ l^+l^-$ (with $l=e$ or $\mu$ or $\tau$) transitions show that these channels can be detected at LHC. The similar  baryonic
 $\Lambda_{b} \rar \Lambda~ \mu^+\mu^-$ has been observed very recently by CDF Collaboration \cite{cdfbiz} and they reported the branching ratio 
of $[1.73\pm0.42($stat$)\pm0.55($syst$)]\times10^{-6}$ which is in  good consistency with our previous work \cite{Azizi3}. Any measurement on the considered channels in the present
work and comparison of the
obtained data with our results can help us  understand better the internal structures of the considered heavy baryons as well as obtain useful information about the distribution
amplitudes of the $\Xi$ and $\Sigma $ baryons. Such comparison in FCNC channels can help us also in the course of searching for new physics effects beyond the SM.

\section{Acknowledgment}
Two of the authors (K. A. and H. S.) would like to thank TUBITAK,
for their partial financial support provided under the project No.110T284.

\newpage

\section*{Appendix A}
In this Appendix,  we present the general decomposition of
the wave functions of the baryons in final states, i.e., $ \epsilon^{abc}\langle 0 | {q_1}_\eta^a(0)
{q_2}_\theta^b(x) {q_3}_\phi^c(0) | B (p)\rangle$
 and   DA's of the $\Xi $ and $\Sigma$ baryons \cite{Yong-LuLiu,Liu}:

\begin{eqnarray}\label{wave func}
&&4\langle0|\epsilon^{abc}{q_1}_\alpha^a(a_1 x){q_2}_\beta^b(a_2
x){q_3}_\gamma^c(a_3 x)|B(p)\rangle\nnb\\
\es\mathcal{S}_1m_{B}C_{\alpha\beta}(\gamma_5B)_{\gamma}+
\mathcal{S}_2m_{B}^2C_{\alpha\beta}(\rlap/x\gamma_5B)_{\gamma}\nnb\\
\ar \mathcal{P}_1m_{B}(\gamma_5C)_{\alpha\beta}B_{\gamma}+
\mathcal{P}_2m_{B}^2(\gamma_5C)_{\alpha\beta}(\rlap/xB)_{\gamma}+
(\mathcal{V}_1+\frac{x^2m_{B}^2}{4}\mathcal{V}_1^M)(\rlap/pC)_{\alpha\beta}(\gamma_5B)_{\gamma}
\nnb\\\ar
\mathcal{V}_2m_{B}(\rlap/pC)_{\alpha\beta}(\rlap/x\gamma_5B)_{\gamma}+
\mathcal{V}_3m_{B}(\gamma_\mu
C)_{\alpha\beta}(\gamma^\mu\gamma_5B)_{\gamma}+
\mathcal{V}_4m_{B}^2(\rlap/xC)_{\alpha\beta}(\gamma_5B)_{\gamma}\nnb\\\ar
\mathcal{V}_5m_{B}^2(\gamma_\mu
C)_{\alpha\beta}(i\sigma^{\mu\nu}x_\nu\gamma_5B)_{\gamma} +
\mathcal{V}_6m_{B}^3(\rlap/xC)_{\alpha\beta}(\rlap/x\gamma_5B)_{\gamma}
+(\mathcal{A}_1\nnb\\
\ar\frac{x^2m_{B}^2}{4}\mathcal{A}_1^M)(\rlap/p\gamma_5
C)_{\alpha\beta}B_{\gamma}+
\mathcal{A}_2m_{B}(\rlap/p\gamma_5C)_{\alpha\beta}(\rlap/xB)_{\gamma}+
\mathcal{A}_3m_{B}(\gamma_\mu\gamma_5 C)_{\alpha\beta}(\gamma^\mu
B)_{\gamma}\nnb\\\ar
\mathcal{A}_4m_{B}^2(\rlap/x\gamma_5C)_{\alpha\beta}B_{\gamma}+
\mathcal{A}_5m_{B}^2(\gamma_\mu\gamma_5
C)_{\alpha\beta}(i\sigma^{\mu\nu}x_\nu B)_{\gamma}+
\mathcal{A}_6m_{B}^3(\rlap/x\gamma_5C)_{\alpha\beta}(\rlap/x
B)_{\gamma}\nnb\\\ar(\mathcal{T}_1+\frac{x^2m_{B}^2}{4}\mathcal{T}_1^M)(p^\nu
i\sigma_{\mu\nu}C)_{\alpha\beta}(\gamma^\mu\gamma_5
B)_{\gamma}+\mathcal{T}_2m_{B}(x^\mu p^\nu
i\sigma_{\mu\nu}C)_{\alpha\beta}(\gamma_5 B)_{\gamma}\nnb\\\ar
\mathcal{T}_3m_{B}(\sigma_{\mu\nu}C)_{\alpha\beta}(\sigma^{\mu\nu}\gamma_5
B)_{\gamma}+
\mathcal{T}_4m_{B}(p^\nu\sigma_{\mu\nu}C)_{\alpha\beta}(\sigma^{\mu\rho}x_\rho\gamma_5
B)_{\gamma}\nnb\\\ar \mathcal{T}_5m_{B}^2(x^\nu
i\sigma_{\mu\nu}C)_{\alpha\beta}(\gamma^\mu\gamma_5 B)_{\gamma}+
\mathcal{T}_6m_{B}^2(x^\mu p^\nu
i\sigma_{\mu\nu}C)_{\alpha\beta}(\rlap/x\gamma_5
B)_{\gamma}\nnb\\
\ar
\mathcal{T}_7m_{B}^2(\sigma_{\mu\nu}C)_{\alpha\beta}(\sigma^{\mu\nu}\rlap/x\gamma_5
B)_{\gamma}+
\mathcal{T}_8m_{B}^3(x^\nu\sigma_{\mu\nu}C)_{\alpha\beta}(\sigma^{\mu\rho}x_\rho\gamma_5
B)_{\gamma}~.\nnb~~~~~~~~~~~~~~~~~~~~~~~(A.1) \end{eqnarray}

The calligraphic functions in the above expression  have not
definite twists but they can be written in terms of the $B$
distribution amplitudes (DA's) with definite and  increasing twists
via   the scalar product $px$.
The relationship between the calligraphic functions appearing in the
above equation and scalar, pseudo-scalar, vector, axial vector and
tensor DA's for B baryon are given in Tables \ref{tab:1}, \ref{tab:2},
\ref{tab:3}, \ref{tab:4} and \ref{tab:5}, respectively.
\begin{table}[h]
\centering
\begin{tabular}{|c|} \hline
$\mathcal{S}_1 = S_1$\\ \hline\hline
 $2px\mathcal{S}_2=S_1-S_2$ \\ \hline
   \end{tabular}
\vspace{0.3cm} \caption{Relations between the calligraphic functions
and B scalar DA's.}\label{tab:1}
\end{table}
\begin{table}[h]
\centering
\begin{tabular}{|c|} \hline
  $\mathcal{P}_1=P_1$\\ \hline
  $2px\mathcal{P}_2=P_1-P_2$ \\ \hline
   \end{tabular}
\vspace{0.3cm} \caption{Relations between the calligraphic functions
and B pseudo-scalar DA's.}\label{tab:2}
\end{table}
\begin{table}[h]
\centering
\begin{tabular}{|c|} \hline
  $\mathcal{V}_1=V_1$ \\ \hline
  $2px\mathcal{V}_2=V_1-V_2-V_3$ \\ \hline
  $2\mathcal{V}_3=V_3$ \\ \hline
  $4px\mathcal{V}_4=-2V_1+V_3+V_4+2V_5$ \\ \hline
  $4px\mathcal{V}_5=V_4-V_3$ \\ \hline
  $4(px)^2\mathcal{V}_6=-V_1+V_2+V_3+V_4
 + V_5-V_6$ \\ \hline
 \end{tabular}
\vspace{0.3cm} \caption{Relations between the calligraphic functions
and B vector DA's.}\label{tab:3}
\end{table}
\begin{table}[h]
\centering
\begin{tabular}{|c|} \hline
  $\mathcal{A}_1=A_1$ \\ \hline
  $2px\mathcal{A}_2=-A_1+A_2-A_3$ \\ \hline
   $2\mathcal{A}_3=A_3$ \\ \hline
  $4px\mathcal{A}_4=-2A_1-A_3-A_4+2A_5$ \\ \hline
  $4px\mathcal{A}_5=A_3-A_4$ \\ \hline
  $4(px)^2\mathcal{A}_6=A_1-A_2+A_3+A_4-A_5+A_6$ \\ \hline
 \end{tabular}
\vspace{0.3cm} \caption{Relations between the calligraphic functions
and B axial vector DA's.}\label{tab:4}
\end{table}
\begin{table}[h]
\centering
\begin{tabular}{|c|} \hline
  $\mathcal{T}_1=T_1$ \\ \hline
  $2px\mathcal{T}_2=T_1+T_2-2T_3$ \\ \hline
  $2\mathcal{T}_3=T_7$ \\ \hline
  $2px\mathcal{T}_4=T_1-T_2-2T_7$ \\ \hline
  $2px\mathcal{T}_5=-T_1+T_5+2T_8$ \\ \hline
  $4(px)^2\mathcal{T}_6=2T_2-2T_3-2T_4+2T_5+2T_7+2T_8$ \\ \hline
  $4px \mathcal{T}_7=T_7-T_8$\\ \hline
  $4(px)^2\mathcal{T}_8=-T_1+T_2 +T_5-T_6+2T_7+2T_8$\\ \hline
 \end{tabular}
\vspace{0.3cm} \caption{Relations between the calligraphic functions
and B tensor DA's.}\label{tab:5}
\end{table}

Every distribution amplitude, $F$=  $S_{1,2}$, $P_{1,2}$, $V_{1\rar6}$,
$A_{1\rar6}$, $T_{1\rar8}$ can be represented as:
\begin{eqnarray}\label{dependent1} F(a_ipx)=\int
dx_1dx_2dx_3\delta(x_1+x_2+x_3-1) e^{-ipx(\sum_{j=1}^3~x_ja_j)}F(x_i)~.\nnb~~~~~~~~~~~~~~~~~~~~~~~~~~~~~~(A.2)
\end{eqnarray}
where, $x_{i}$ with $i=1,~2$ or $3$ are longitudinal momentum
fractions carried by the participating quarks.

The explicit expressions for the   DA's of the $B$ baryon up to twists six are
given as \cite{Yong-LuLiu,Liu}: \\
Twist-$3$ distribution
amplitudes:
\begin{eqnarray}
V_1(x_i)&=&120x_1x_2x_3\phi_3^0\,,\hspace{2.5cm}A_1(x_i)=0\,,\nonumber\\
T_1(x_i)&=&120x_1x_2x_3\phi_3^{'0}\,.
\end{eqnarray}
Twist-$4$ distribution amplitudes:
\begin{eqnarray}
S_1(x_i)&=&6(x_2-x_1)x_3(\xi_4^0+\xi_4^{'0})\,,
\hspace{1.6cm}P_1(x_i)=6(x_2-x_1)x_3(\xi_4^0-\xi_4^{'0})\,,\nonumber\\
V_2(x_i)&=&24x_1x_2\phi_4^0\,,\hspace{3.9cm}A_2(x_i)=0\,,\nonumber\\
V_3(x_i)&=&12x_3(1-x_3)\psi_4^0\,,
\hspace{2.8cm}A_3(x_i)=-12x_3(x_1-x_2)\psi_4^0\,,\nonumber\\
T_2(x_i)&=&24x_1x_2\phi_4^{'0}\,,
\hspace{3.9cm}T_3(x_i)=6x_3(1-x_3)(\xi_4^0+\xi_4^{'0})\,,\nonumber\\
T_7(x_i)&=&6x_3(1-x_3)(\xi_4^{'0}-\xi_4^0)\,.
\end{eqnarray}
Twist-$5$ distribution amplitudes:
\begin{eqnarray}
S_2(x_i)&=&\frac32(x_1-x_2)(\xi_5^0+\xi_5^{'0})\,,
\hspace{1.5cm}P_2(x_i)=\frac32(x_1-x_2)(\xi_5^0-\xi_5^{'0})\,,\nonumber\\
V_4(x_i)&=&3(1-x_3)\psi_5^0\,,
\hspace{2.95cm}A_4(x_i)=3(x_1-x_2)\psi_5^0\,,\nonumber\\
V_5(x_i)&=&6x_3\phi_5^0\,,\hspace{4.05cm}A_5(x_i)=0\,,\nonumber\\
T_4(x_i)&=&-\frac32(x_1+x_2)(\xi_5^{'0}+\xi_5^0)\,,
\hspace{1.25cm}T_5(x_i)=6x_3\phi_5^{'0}\,,\nonumber\\
T_8(x_i)&=&\frac32(x_1+x_2)(\xi_5^{'0}-\xi_5^0)\,.
\end{eqnarray}
Twist-$6$ distribution amplitudes:
\begin{eqnarray}
V_6(x_i)&=&2\phi_6^0\,,\hspace{2.5cm}A_6(x_i)=0\,,\nonumber\\
T_6(x_i)&=&2\phi_6^{'0}\,.
\end{eqnarray}
where,
\begin{eqnarray}
\phi_3^0&=&\phi_6^0=f_{B},
\hspace{2.8cm}\psi_4^0=\psi_5^0=\frac12(f_{B}-\lambda_1)\,,\nonumber\\
\phi_4^0&=&\phi_5^0=\frac12(f_{B}+\lambda_1),
\hspace{1.3cm}\phi_3'^0=\phi_6'^0=-\xi_5^0=\frac16(4\lambda_3-\lambda_2)\,,\nonumber\\
\phi_4'^0&=&\xi_4^0=\frac16(8\lambda_3-3\lambda_2),
\hspace{1.2cm}\phi_5'^0=-\xi_5'^0=\frac16\lambda_2\,,\nonumber\\
\xi_4'^0&=&\frac16(12\lambda_3-5\lambda_2)\,.
\end{eqnarray}

\begin{table}[h]
\renewcommand{\arraystretch}{1.5}
\addtolength{\arraycolsep}{3pt}
$$
\begin{array}{|c|c|c|c|c|}

\hline \hline
                & \mbox{a} & \mbox{b}  & m_{fit}& q^2=0
                \\
\hline
 f_1            &   0.166  &  -0.024  &  5.35  &   0.142\pm  0.036  \\
 f_2            &   0.028  &  -0.048  &  5.31  &  -0.020 \pm  0.005  \\
 f_3            &  -0.004  &  -0.006   &  5.37  &  -0.010 \pm  0.002 \\
 g_1            &   0.106   &  0.054  &  5.24  &   0.160 \pm 0.042  \\
 g_2            &  -0.005  &  -0.004   &  5.28  &  -0.009 \pm  0.002 \\
 g_3            &   0.003  &  -0.006   &  4.70  &  -0.003  \pm 0.001 \\
 f_1^{T}        &   0.127  &  -0.129  &  5.10  &  -0.0020  \pm  0.0005 \\
 f_2^{T}        &   0.072  &   0.085  &  5.40  &   0.157  \pm  0.041\\
 f_3^{T}        &  -0.003  &   0.049  &  5.23  &   0.046  \pm  0.011\\
 g_1^{T}        &   0.288  &  -0.312   &  4.80  &  -0.024 \pm  0.006 \\
 g_2^{T}        &   0.036  &   0.119  &  4.70  &  0.155   \pm  0.040\\
 g_3^{T}        &   0.024  &  -0.095  &  5.33  &  -0.071  \pm  0.018\\
\hline \hline
\end{array}
$$
\caption{Parameters appearing in  the fit function of the  form
factors and the values of the form factors at $q^2=0$  for
$\Xi_{b}\rightarrow \Xi\ell^{+}\ell^{-}$. } \label{tab:13}
\renewcommand{\arraystretch}{1}
\addtolength{\arraycolsep}{-1.0pt}
\end{table}

\begin{table}[h]
\renewcommand{\arraystretch}{1.5}
\addtolength{\arraycolsep}{3pt}
$$
\begin{array}{|c|c|c|c|c|}

\hline \hline
                & \mbox{a} & \mbox{b}  & m_{fit}& q^2=0
                \\
\hline
 f_1            &   0.035  &   0.011  &  5.16  &   0.046  \pm 0.011  \\
 f_2            &   0.027  &  -0.060  &  5.32  &  -0.033  \pm  0.008 \\
 f_3            &   0.086  &  -0.110  &  5.38  &  -0.024  \pm  0.006  \\
 g_1            &   0.047  &   0.020  &  5.34  &   0.067  \pm  0.017 \\
 g_2            &  -0.003   &  -0.021  &  5.25  &  -0.024 \pm  0.006 \\
 g_3            &  -0.003  &  -0.024  &  5.39  &  -0.027  \pm  0.006 \\
 f_1^{T}        &   0.045  &  -0.047  &  5.29  &  -0.0020  \pm  0.0005 \\
 f_2^{T}        &   0.034  &   0.015  &  5.25  &   0.049 \pm   0.012 \\
 f_3^{T}        &  -0.145  &   0.168  &  5.17  &  0.023  \pm   0.006\\
 g_1^{T}        &   0.006  &  -0.012  &  4.67  &  -0.006  \pm  0.001 \\
 g_2^{T}        &  -0.041  &   0.054  &  5.38  &   0.013  \pm  0.003 \\
 g_3^{T}        &   0.049  &  -0.071   &  5.36  &  -0.022 \pm  0.005  \\
\hline \hline
\end{array}
$$
\caption{Parameters appearing in  the fit function of the  form
factors and the values of the form factors at $q^2=0$  for
$\Xi_{b}\rightarrow \Sigma\ell^{+}\ell^{-}$.} \label{tab:14}
\renewcommand{\arraystretch}{1}
\addtolength{\arraycolsep}{-1.0pt}
\end{table}

%%%%%%%%%%%%%%%%%%%%%%%%%%%%%%%%%%%%%%%%%%%%%%%%%%
\begin{table}[h]
\renewcommand{\arraystretch}{1.5}
\addtolength{\arraycolsep}{3pt}
$$
\begin{array}{|c|c|c|c|c|}

\hline \hline
                & \mbox{a} & \mbox{b}  & m_{fit}& q^2=0
                \\
\hline
 f_1            & 0.526    &-0.116    &1.53    &0.409 \pm   0.106    \\
 f_2            & -0.550    &0.026    &1.58    &-0.524 \pm   0.136  \\
 f_3            & -0.204    &-0.582    &1.57    &-0.786  \pm  0.204\\
 g_1            & 0.183    &0.154    &1.55    &0.337  \pm  0.088\\
 g_2            &-0.431     &0.045    &1.63    &-0.386  \pm0.100   \\
 g_3            &-0.190     &-0.285    &1.63    &-0.475  \pm 0.123  \\
 f_1^{T}         & 0.042   &-0.048    &1.56    &-0.006  \pm0.001   \\
 f_2^{T}         &0.585     &-0.125    &1.52    &0.460  \pm0.120   \\
 f_3^{T}         &-0.449     &1.127    &1.59    &0.678  \pm0.176   \\
 g_1^{T}         &0.058     &-0.062    &1.58    &-0.004   \pm0.001 \\
 g_2^{T}         &0.730     &-0.201    &1.57    &0.529   \pm0.260  \\
 g_3^{T}        &-0.531     &-0.148    &1.61    &-0.679   \pm0.176  \\
\hline \hline
\end{array}
$$
\caption{Parameters appearing in  the fit function of the  form
factors and the values of the form factors at $q^2=0$  for
$\Xi_{c}\rightarrow \Sigma\ell^{+}\ell^{-}$.} \label{tab:1444}
\renewcommand{\arraystretch}{1}
\addtolength{\arraycolsep}{-1.0pt}
\end{table}
%%%%%%%%%%%%%%%%%%%%%%%%%%%%%%%%%%%%%%%%%%%%%%%%%%
\begin{table}[h]
\renewcommand{\arraystretch}{1.5}
\addtolength{\arraycolsep}{3pt}
$$
\begin{array}{|c|c|c|c|c|}

\hline \hline
                & \mbox{a} & \mbox{b}  & m_{fit}& q^2=0
                \\
\hline
 f_1            &   0.092  &  -0.003  &  5.30  &   0.089 \pm 0.022  \\
 f_2            &  -0.010  &  -0.021  &  5.32  &  -0.031 \pm 0.007  \\
 f_3            &   0.015  &  -0.058  &  5.73  &  -0.043 \pm  0.010 \\
 g_1            &  -0.421  &   0.477  &  5.20  &   0.056  \pm 0.014   \\
 g_2            &  -0.012  &  -0.008  &  5.10  &  -0.020  \pm  0.005 \\
 g_3            &  -0.035  &   0.001  &  5.00  &  -0.034  \pm  0.008\\
 f_1^{T}        &  -1.126  &   1.124   &  5.40  &  -0.0020 \pm  0.0005 \\
 f_2^{T}        &   0.028  &   0.081  &  4.80  &   0.109  \pm 0.028 \\
 f_3^{T}        &   0.035  &   0.132  &  5.26  &   0.167  \pm 0.043 \\
 g_1^{T}        &   0.645  &  -0.645  &  5.40  &  0.000   \pm 0.000\\
 g_2^{T}        &   0.022  &   0.002  &  4.80  &   0.024  \pm 0.006 \\
 g_3^{T}        &  -0.210  &  -0.058  &  5.32  &  -0.268   \pm 0.070\\
\hline \hline
\end{array}
$$
\caption{Parameters appearing in  the fit function of the  form
factors and the values of the form factors at $q^2=0$  for
$\Xi'_{b}\rightarrow \Xi\ell^{+}\ell^{-}$.} \label{tab:15}
\renewcommand{\arraystretch}{1}
\addtolength{\arraycolsep}{-1.0pt}
\end{table}

\begin{table}[h]
\renewcommand{\arraystretch}{1.5}
\addtolength{\arraycolsep}{3pt}
$$
\begin{array}{|c|c|c|c|c|}

\hline \hline
                & \mbox{a} & \mbox{b}  & m_{fit}& q^2=0
                \\
\hline
 f_1            &   0.034  &   0.056  &  5.13  &   0.090  \pm 0.022 \\
 f_2            &   0.046  &  -0.104  &  5.34  &  -0.058   \pm 0.014\\
 f_3            &   0.055  &  -0.104  &  5.27  &  -0.049   \pm 0.012\\
 g_1            &  -0.237  &   0.275  &  5.36  &   0.038   \pm  0.010\\
 g_2            &   0.008  &  -0.049  &  5.34  &  -0.041   \pm 0.011\\
 g_3            &  -0.006  &  -0.039  &  5.31  &  -0.045   \pm 0.011\\
 f_1^{T}        &   0.458  &  -0.458    &  5.15  &  0.000 \pm  0.000 \\
 f_2^{T}        &  -0.541  &   0.679  &  5.35  &   0.138  \pm  0.036\\
 f_3^{T}        &  -0.281  &   0.494  &  5.38  &   0.213  \pm  0.055\\
 g_1^{T}        &   0.722  &  -0.725  &  5.08  &  -0.003  \pm  0.001\\
 g_2^{T}        &  -0.106  &   0.191  &  5.28  &   0.085   \pm 0.021\\
 g_3^{T}        &  0.025   &  -0.327   &  5.32  &  -0.302   \pm 0.078\\
\hline \hline
\end{array}
$$
\caption{Parameters appearing in  the fit function of the  form
factors and the values of the form factors at $q^2=0$  for
$\Xi^{'}_{b}\rightarrow \Sigma\ell^{+}\ell^{-}$.} \label{tab:151}
\renewcommand{\arraystretch}{1}
\addtolength{\arraycolsep}{-1.0pt}
\end{table}

\begin{table}[h]
\renewcommand{\arraystretch}{1.5}
\addtolength{\arraycolsep}{3pt}
$$
\begin{array}{|c|c|c|c|c|}

\hline \hline
                & \mbox{a} & \mbox{b}  & m_{fit}& q^2=0
                \\
\hline
 f_1            &   -0.564  &   0.640  &  1.52  &   0.076 \pm 0.019  \\
 f_2            &   -0.426  &  -0.258  &  1.55  &  -0.684 \pm  0.178 \\
 f_3            &   -0.642  &  -0.297  &  1.58  &  -0.939  \pm  0.244\\
 g_1            &   -0.092  &  0.212   &  1.62  &   0.120  \pm  0.031\\
 g_2            &   -0.265  &  -0.081  &  1.60  &  -0.346 \pm   0.090\\
 g_3            &    0.238  &  -0.349  &  1.55  &  -0.111 \pm   0.029\\
 f_1^{T}        &    0.272  &   -0.293 &  1.60  &  -0.021 \pm   0.005\\
 f_2^{T}        &    0.432  &   0.112  &  1.53  &   0.544 \pm   0.141\\
 f_3^{T}        &   -0.433  &   0.605  &  1.62  &   0.172 \pm   0.045\\
 g_1^{T}        &    0.258  &  -0.265  &  1.72  &  -0.007 \pm   0.002\\
 g_2^{T}        &    0.401  &  -0.013  &  1.50  &  0.388  \pm  0.101\\
 g_3^{T}        &    0.153  &  -0.510   &  1.63  &  -0.357 \pm   0.093\\
\hline \hline
\end{array}
$$
\caption{Parameters appearing in  the fit function of the  form
factors and the values of the form factors at $q^2=0$  for
$\Xi^{'}_{c}\rightarrow \Sigma\ell^{+}\ell^{-}$.} \label{tab:152}
\renewcommand{\arraystretch}{1}
\addtolength{\arraycolsep}{-1.0pt}
\end{table}

\begin{table}[h]
\renewcommand{\arraystretch}{1.5}
\addtolength{\arraycolsep}{3pt}
$$
\begin{array}{|c|c|c|c|c|}

\hline \hline
                & \mbox{a} & \mbox{b}  & m_{fit}& q^2=0
                \\
\hline
 f_1            &  -0.4142  &   0.608  &  1.52  &   0.194  \pm 0.050  \\
 f_2            &  -0.320  &  -0.036  &  1.60  &  -0.356 \pm  0.092  \\
 f_3            &  1.068   &  -1.530  &  1.55  &  -0.462  \pm  0.120\\
 g_1            &  -0.624  &   0.935  &  1.58  &   0.311  \pm  0.081\\
 g_2            &   0.010 & - 0.161  &  1.63  &  -0.151  \pm  0.038\\
 g_3            &  1.398   & - 1.710  &  1.61  &  -0.312 \pm   0.081\\
 \hline \hline
\end{array}
$$
\caption{Parameters appearing in  the fit function of the  form
factors and the values of the form factors at $q^2=0$  for
$\Xi_{c}\rightarrow \Xi\ell \nu$.} \label{tab:16}
\renewcommand{\arraystretch}{1}
\addtolength{\arraycolsep}{-1.0pt}
\end{table}

\begin{table}[h]
\renewcommand{\arraystretch}{1.5}
\addtolength{\arraycolsep}{3pt}
$$
\begin{array}{|c|c|c|c|c|}

\hline \hline
                & \mbox{a} & \mbox{b}  & m_{fit}& q^2=0
                \\
\hline
 f_1            &   0.528  & -0.119  &  1.52  &   0.409  \pm 0.106\\
 f_2            &  -0.564  &  0.006  &  1.58  &  -0.558  \pm 0.145 \\
 f_3            &  -0.598  & -0.133  &  1.57  &  -0.731   \pm 0.190\\
 g_1            &   0.448  & -0.031  &  1.55  &   0.417   \pm 0.104\\
 g_2            &  -0.493  &  0.096  &  1.61  &  -0.397   \pm 0.099\\
 g_3            &  -0.295  & -0.329  &  1.63  &  -0.624   \pm 0.150\\
 \hline \hline
\end{array}
$$
\caption{Parameters appearing in  the fit function of the  form
factors and the values of the form factors at $q^2=0$  for
$\Xi_{c}\rightarrow \Sigma\ell \nu$.} \label{tab:17}
\renewcommand{\arraystretch}{1}
\addtolength{\arraycolsep}{-1.0pt}
\end{table}

\begin{table}[h]
\renewcommand{\arraystretch}{1.5}
\addtolength{\arraycolsep}{3pt}
$$
\begin{array}{|c|c|c|c|c|}

\hline \hline
                & \mbox{a} & \mbox{b}  & m_{fit}& q^2=0
                \\
\hline
 f_1            &  -1.498  &   2.075  &  1.60  &   0.577   \pm 0.150\\
 f_2            &  -0.359  &  -0.142  &  1.66  &  -0.501  \pm 0.130 \\
 f_3            &  -0.760  &   0.082  &  1.70  &  -0.678   \pm 0.176\\
 g_1            &   0.159  &   0.292  &  1.57  &   0.451    \pm0.113\\
 g_2            &  -0.317  &  -0.024  &  1.62  &  -0.341    \pm0.089\\
 g_3            &   0.976  &  -1.218  &  1.62  &  -0.242   \pm 0.061\\
 \hline \hline
\end{array}
$$
\caption{Parameters appearing in  the fit function of the  form
factors and the values of the form factors at $q^2=0$  for
$\Xi'_{c}\rightarrow \Xi\ell \nu$.} \label{tab:18}
\renewcommand{\arraystretch}{1}
\addtolength{\arraycolsep}{-1.0pt}
\end{table}

\begin{table}[h]
\renewcommand{\arraystretch}{1.5}
\addtolength{\arraycolsep}{3pt}
$$
\begin{array}{|c|c|c|c|c|}

\hline \hline
                & \mbox{a} & \mbox{b}  & m_{fit}& q^2=0
                \\
\hline
 f_1            &  -0.564  &   0.640  &  1.52  &   0.076  \pm 0.020 \\
 f_2            &  -0.226  &  -0.427  &  1.55  &  -0.653  \pm  0.169\\
 f_3            &  -1.007  &   0.112  &  1.58  &  -0.895   \pm 0.232\\
 g_1            &  -0.017  &   0.054  &  1.62  &   0.037   \pm 0.009\\
 g_2            &  -0.265  &  -0.081  &  1.60  &  -0.346   \pm 0.089\\
 g_3            &  0.238   &  -0.349  &  1.55  &  -0.111   \pm 0.028\\
 \hline \hline
\end{array}
$$
\caption{Parameters appearing in  the fit function of the  form
factors and the values of the form factors at $q^2=0$  for
$\Xi^{'}_{c}\rightarrow \Sigma\ell \nu$.} \label{tab:181}
\renewcommand{\arraystretch}{1}
\addtolength{\arraycolsep}{-1.0pt}
\end{table}

\begin{table}[h]\centering
$$
\begin{array}{|c|c|c|c|}

\hline \hline
    & \Gamma(GeV)   \\

\hline
 \Xi_{b} \rar \Xi e^+e^- & (9.941 \pm 2.982) \times
10^{-19}
\\ \hline
\Xi_{b} \rar \Xi \mu^+\mu^- & (9.872 \pm 3.455) \times 10^{-19}
\\ \hline
\Xi_{b}\rar \Xi \tau^+\tau^- & (1.611 \pm 0.483) \times 10^{-19}
\\ \hline
\Xi_{b} \rar \Sigma e^+e^- & (6.359 \pm 2.225) \times 10^{-20}
\\ \hline
\Xi_{b} \rar \Sigma \mu^+\mu^- & (6.194 \pm 2.167) \times 10^{-20}
\\ \hline
\Xi_{b} \rar \Sigma \tau^+\tau^- & (4.338 \pm 1.518) \times 10^{-20}
\\ \hline
\Xi_{c} \rar \Sigma e^+e^- & (2.162 \pm 0.757) \times 10^{-25}
\\ \hline
\Xi_{c} \rar \Sigma \mu^+\mu^- & (2.152 \pm 0.753 ) \times 10^{-25}
\\ \hline
\Xi_{c} \rar \Xi e^+\nu_{e} & (4.264 \pm 1.49) \times 10^{-13}
\\ \hline
\Xi_{c} \rar \Xi \mu^+\nu_{\mu} & ( 4.202\pm 1.26) \times 10^{-13}
\\ \hline
\Xi_{c} \rar \Sigma e^+\nu_{e} & (2.204 \pm 0.771) \times 10^{-14}
\\ \hline
\Xi_{c} \rar \Sigma \mu^+\nu_{\mu} & (2.183 \pm 0.764) \times
10^{-14}
\\ \hline
 \Xi'_{b}\rightarrow \Xi e^{+}e^{-}&(2.411 \pm 0.843) \times
10^{-17}
\\ \hline
 \Xi'_{b}\rightarrow \Xi \mu^{+}\mu^{-}&(2.407 \pm 0.842) \times
10^{-17}
\\ \hline
 \Xi'_{b}\rightarrow \Xi \tau^{+}\tau^{-}&(1.199 \pm 0.419) \times
10^{-17}
\\ \hline
\Xi'_{c}\rightarrow \Xi \mu^{+}\nu_{\mu}&(1.009 \pm 0.303) \times
10^{-12}
\\ \hline
\Xi'_{c}\rightarrow \Xi e^{+}\nu_{e}&(1.109 \pm 0.388) \times
10^{-12}
\\ \hline
\Xi'_{c}\rightarrow \Sigma e^{+}\nu_{e}&(8.425 \pm 2.948) \times
10^{-14}
\\ \hline
\Xi'_{c}\rightarrow \Sigma \mu^{+}\nu_{\mu}&(8.340 \pm 2.919) \times
10^{-14}
\\ \hline
\Xi'_{c}\rightarrow \Sigma e^{+}e^{-}&(1.718 \pm 0.515) \times
10^{-24}
\\ \hline
\Xi'_{c}\rightarrow \Sigma \mu^{+}\mu^{-}&(1.666 \pm 0.499) \times
10^{-24}
\\ \hline
\Xi'_{b}\rightarrow \Sigma e^{+}e^{-}&(3.815 \pm 1.335) \times
10^{-19}
\\ \hline
\Xi'_{b}\rightarrow \Sigma \mu^{+}\mu^{-}&(3.813 \pm 1.334) \times
10^{-19}
\\ \hline
\Xi'_{b}\rightarrow \Sigma \tau^{+}\tau^{-}&(2.783 \pm 0.974) \times
10^{-19}
\\ \hline
\end{array}
$$
\vspace{0.8cm} \caption{The values of the decay rates  in full
theory for different leptons.} \label{tab:27}
\end{table}

\begin{table}[h]\centering
$$
\begin{array}{|c|c|c|c|}

\hline \hline
    & BR  \\

\hline
 \Xi_{b} \rar \Xi e^+e^- & (2.25 \pm 0.78) \times
10^{-6}
\\ \hline
\Xi_{b} \rar \Xi \mu^+\mu^- & (2.23 \pm 0.67) \times 10^{-6}
\\ \hline
\Xi_{b}\rar \Xi \tau^+\tau^- & (0.36 \pm 0.11) \times 10^{-6}
\\ \hline
\Xi_{b} \rar \Sigma e^+e^- & (1.44 \pm 0.50) \times 10^{-7}
\\ \hline
\Xi_{b} \rar \Sigma \mu^+\mu^- & (1.40 \pm 0.49) \times 10^{-7}
\\ \hline
\Xi_{b} \rar \Sigma \tau^+\tau^- & (0.98 \pm 0.29) \times 10^{-7}
\\ \hline
\Xi_{c} \rar \Sigma e^+e^- & ( 3.68\pm 1.29) \times 10^{-14}
\\ \hline
\Xi_{c} \rar \Sigma \mu^+\mu^- & (3.66 \pm 1.28 ) \times 10^{-14}
\\ \hline
\Xi_{c} \rar \Xi e^+\nu_{e} & (7.26 \pm 2.54 ) \times 10^{-2}
\\ \hline
\Xi_{c} \rar \Xi \mu^+\nu_{\mu} & (7.15 \pm 2.50) \times 10^{-2}
\\ \hline
\Xi_{c} \rar \Sigma e^+\nu_{e} & (1.48 \pm 0.52) \times 10^{-2}
\\ \hline
\Xi_{c} \rar \Sigma \mu^+\nu_{\mu} & (1.47 \pm 0.51) \times 10^{-2}
\\ \hline
\end{array}
$$
\vspace{0.8cm} \caption{The values of the branching ratios  in full
theory for different leptons.} \label{tab:27777}
\end{table}

\end{document}